
\input harvmac.tex      
\input epsf.tex         
\def\sc{\scriptstyle}
\def\tr#1{{\rm Tr}_{#1}}
\def\mass{{\rm {\hat{m}} }}
\def\ma{\mass}
\def\ha{ {\scriptstyle{\inv{2}} }}
\def\mtoo{{\mathop{\longrightarrow}^{\ma \to 0}}}
\def\dag{\dagger}
\def\betah{{\hat{\beta}}}
\def\hp{\CH_P}
\def\hf{\CH_F}
\def\qsl{{q-\hat{sl(2)}}}
\def\slh{{\hat{sl(2)}}}
\def\psib{\bar{\psi}}
\def\psip{\psi_+}
\def\psim{\psi_-}
\def\psibp{\psib_+}
\def\psibm{\psib_-}

\def\vphi{\varphi}
\def\bh{\hat{b}}
\def\bhp{\bh^+}
\def\bhm{\bh^-}
\def\sqm{\sqrt{\ma}}
\def\ez{e^{\ma z u + \ma \zb /u}}
\def\squ{\sqrt{u}}
\def\bp{b^+}
\def\bm{b^-}
\def\bb{\bar{b}}
\def\bbp{\bar{b}^+}
\def\bbm{\bar{b}^-}
\def\du{ \frac{du}{2 \pi i |u|} }
\def\dua{ \frac{du}{2 \pi i u} }
\def\Qt{\tilde{Q}}

\def\Qpm#1{ Q^\pm_#1 }

\def\mn{\frac{ \ma^{|n|+|m|} }{\ma^{|n+m|} } }

\def\slhh{\hat{\hat{sl(2)}}}
\def\chih{\hat{\chi}}
\def\chihh{\hat{\chih}}
\def\bbull{b_\bullet}
\def\bbul{\bbull}
\def\bbbul{\bb_\bullet}

\def\lvacp{\langle +\ha \vert}
\def\lvacm{\langle -\ha \vert}
\def\lvacmp{\langle \mp \ha \vert}
\def\lvacpm{\langle \pm \ha \vert}
\def\rvacpm{\vert \pm \ha \rangle}
\def\rvacmp{\vert \mp \ha \rangle}
\def\rvacp{\vert + \ha \rangle}
\def\rvacm{\vert - \ha \rangle}
\def\va#1{\vert {#1} \rangle}
\def\lva#1{\langle {#1} \vert}

\def\mat#1#2#3#4{\left(\matrix{#1&#2\cr #3 &#4 \cr}\right)}
\def\col#1#2{\left(\matrix{#1\cr #2\cr}\right)}
\def\dz{\frac{dz}{2\pi i}}
\def\dzb{\frac{d\zb}{2\pi i}}
\def\dzf{\frac{dz}{4\pi i}}
\def\dzbf{\frac{d\zb}{4\pi i}}

\def\uone{U(1)}
\def\psipm{\psi_\pm}
\def\psibpm{\psib_\pm }
\def\bhpm{\bh^\pm}
\def\bpm{b^\pm}
\def\bbpm{\bar{b}^\pm}
\def\bmp{b^\mp}
\def\bbmp{\bar{b}^\mp}
\def\denm{\de_{n, -m} }
\def\bhpm{\bh^\pm}
\def\cL{\CC^{L}_{\vphi}}
\def\cR{\CC^{R}_{\vphi} }

\def\hal{{\CH^L_a}}
\def\har{{\CH^R_a}}

\def\halp{{\CH^L_{a_+}}}
\def\halm{{\CH^L_{a_-}}}

\def\vh{\hat{V}}
\def\vhh{\hat{\hat{V}}}

%
%
%

\def\tilde{\widetilde}
\def\bar{\overline}
\def\hat{\widehat}
\def\*{\star}
\def\[{\left[}
\def\]{\right]}
\def\({\left(}		
\def\){\right)}

%
%
\def\zb{{\bar{z} }}
\def\frac#1#2{{#1 \over #2}}
\def\inv#1{{1 \over #1}}

\def\d{\partial}

\def\rvac{\hbox{$\vert 0\rangle$}}
\def\lvac{\hbox{$\langle 0 \vert $}}
\def\2pi{\hbox{$2\pi i$}}
\def\e#1{{\rm e}^{^{\textstyle #1}}}

\def\dsl{\raise.15ex\hbox{/}\kern-.57em\partial}
\def\Dsl{\,\raise.15ex\hbox{/}\mkern-.13.5mu D}
%
%
\def\th{\theta}		
		\def\Ga{\Gamma}

\def\al{\alpha}
\def\ep{\epsilon}
\def\la{\lambda}	\def\La{\Lambda}
\def\de{\delta}		\def\De{\Delta}
\def\om{\omega}		
	
\def\vphi{\varphi}
%
%
		\def\CC{{\cal C}}
		
	\def\CH{{\cal H}}	
		
		\def\CO{{\cal O}}

\def\rvac{\hbox{$\vert 0\rangle$}}
\def\lvac{\hbox{$\langle 0 \vert $}}

\def\2pi{\hbox{$2\pi i$}}
\def\e#1{{\rm e}^{^{\textstyle #1}}}

\def\dsl{\raise.15ex\hbox{/}\kern-.57em\partial}
\def\Dsl{\,\raise.15ex\hbox{/}\mkern-.13.5mu D}
%
%
%
\font\numbers=cmss12
\font\upright=cmu10 scaled\magstep1
\def\stroke{\vrule height8pt width0.4pt depth-0.1pt}
\def\topfleck{\vrule height8pt width0.5pt depth-5.9pt}
\def\botfleck{\vrule height2pt width0.5pt depth0.1pt}
\def\Zmath{\vcenter{\hbox{\numbers\rlap{\rlap{Z}\kern
0.8pt\topfleck}\kern
2.2pt
                   \rlap Z\kern 6pt\botfleck\kern 1pt}}}
\def\Qmath{\vcenter{\hbox{\upright\rlap{\rlap{Q}\kern
                   3.8pt\stroke}\phantom{Q}}}}
\def\Nmath{\vcenter{\hbox{\upright\rlap{I}\kern 1.7pt N}}}
\def\Cmath{\vcenter{\hbox{\upright\rlap{\rlap{C}\kern
                   3.8pt\stroke}\phantom{C}}}}
\def\Rmath{\vcenter{\hbox{\upright\rlap{I}\kern 1.7pt R}}}
\def\Z{\ifmmode\Zmath\else$\Zmath$\fi}
\def\Q{\ifmmode\Qmath\else$\Qmath$\fi}
\def\N{\ifmmode\Nmath\else$\Nmath$\fi}
\def\C{\ifmmode\Cmath\else$\Cmath$\fi}
\def\R{\ifmmode\Rmath\else$\Rmath$\fi}
\def\Zmath{Z}


\def\FF{R. Fehlmann and G. Felder, private communication, unpublished.}
\def\AASS{E. Abdalla, M. C. B. Abdalla, G. Sotkov, and M. Stanishkov,
{\it Off Critical Current Algebras}, Univ. Sao Paulo preprint,
IFUSP-preprint-1027, Jan. 1993.}
\def\Ginsparg{P. Ginsparg, Les Houches Lectures 1988, in
{\it Fields, Strings and Critical Phenomena}, E. Br\'ezin and J. Zinn-Justin
Eds., North Holland (1990).}
\def\Griffin{P. Griffin, Nucl. Phys. B334, 637.}
\def\MSS{B. Schroer and T. T. Truong, Nucl. Phys. B144 (1978) 80 \semi
E. C. Marino, B. Schroer, and J. A. Swieca, Nucl. Phys. B200 (1982) 473.}

\def\VVstar{B. Davies, O. Foda, M. Jimbo, T. Miwa and A. Nakayashiki,
Commun. Math. Phys. 151 (1993) 89;
M. Jimbo, K. Miki, T. Miwa and A. Nakayashiki, {\it Correlation
Functions of the XXZ model for $\Delta < -1$}, Kyoto 1992  preprint,
PRINT-92-0101,
hep-th/9205055.}
\def\Raja{R. K. Kaul and R. Rajaraman, {\it Non-Local Currents in the Massive
Thirring Model}, Indian Inst. of Sci. preprint  CTS 5-92, 1992.}
\def\Rajab{S.-J. Chang and R. Rajaraman, {\it New Non-Local Currents in the
Quantum Sine-Gordon Model}, Univ. of Illinois at Urbana-Champaign preprint
1993.}
\def\BLnlc{D. Bernard and A. LeClair, Commun. Math. Phys. 142 (1991) 99;
Phys. Lett. B247 (1990) 309.}
\def\form{F. A. Smirnov, {\it Form Factors in Completely Integrable
Models of Quantum Field Theory}, in {\it Advanced Series in Mathematical
Physics} 14, World Scientific, 1992.}
\def\Mand{S. Mandelstam, Phys. Rev. D11 (1975) 3026.}
\def\BPZ{A. A. Belavin, A. M. Polyakov, and A. B. Zamolodchikov,
Nucl. Phys. B241 (1984) 333.}
\def\KZ{V. G. Knizhnik and A. B. Zamolodchikov, Nucl. Phys. B247 (1984) 83.}
\def\Drinfeld{V. G. Drinfel'd, Sov. Math. Dokl. 32 (1985) 254;
Sov. Math. Dokl. 36 (1988) 212. }
\def\Jimbo{M. Jimbo, Lett. Math. Phys. 10 (1985) 63; Lett. Math. Phys. 11
(1986) 247; Commun. Math. Phys. 102 (1986) 537.}
\def\STF{E. K. Sklyanin, L. A. Takhtadzhyan, and L. D. Faddeev, Theor.
Math. 40 (1980) 688; L. Faddeev, Les Houches Lectures 1982, Elsevier
(1984).}
\def\GO{P. Goddard and D. Olive,
Int. Jour. Mod. Phys. A 1 (1986) 303.}
\def\Kac{V. G. Kac, {\it Infinite Dimensional Lie Algebras}, Cambridge
University Press (1985).}
\def\ZZ{A. B. Zamolodchikov and Al. B. Zamolodchikov, Annals
Phys. 120 (1979) 253.}
\def\Colemani{S. Coleman, Phys. Rev. D 11 (1975) 2088.}
\def\TI{H. Itoyama and H. B. Thacker, Nucl. Phys. B320 (1989) 541.}
\def\Zamoiii{A. B. Zamolodchikov, Int. Journ. of Mod. Phys. A4 (1989) 4235;
in Adv. Studies in Pure Math. vol. 19 (1989) 641.}
%

\Title{CLNS 93/1220, hep-th/9305110 (revised) }
{\vbox{\centerline{Spectrum Generating Affine Lie Algebras    }
\centerline{ in Massive Field Theory} }}

\bigskip
\bigskip

\centerline{Andr\'e LeClair}
\medskip\centerline{Newman Laboratory}
\centerline{Cornell University}
\centerline{Ithaca, NY  14853}
\bigskip\bigskip

\vskip .3in

We present a new application of affine Lie algebras to
massive quantum field theory in 2 dimensions, by investigating
the $q\to 1$ limit of the q-deformed affine $\hat{sl(2)}$  symmetry of
the sine-Gordon theory, this limit occurring at the free fermion point.
We describe how radial quantization leads to a quasi-chiral factorization of
the space of fields.
The
conserved charges which generate the
affine Lie algebra split into two independent affine algebras on
this factorized space, each with level 1, in the anti-periodic sector.
The space of fields in the anti-periodic sector can be organized
using level-$1$ highest weight representations, if one supplements
the $\slh$ algebra with the usual local integrals of motion.
Using the integrals of motion, a momentum
space bosonization involving vertex operators is formulated.
This leads to a new way of computing form-factors, as
vacuum expectation values  in momentum space.

\Date{10/93}
\draftmode
%
%
%
%
%
\noblackbox
\def\ot{\otimes}
\def\zb{{\bar{z}}}

\def\zbar{{\bar{z}}}

%
%
%
%
%
%
%
%
%
%

\newsec{Introduction}

Algebraic methods are becoming increasingly important for solving
quantum field theory non-perturbatively.  This is especially true
for the massless conformal field theories in two spacetime dimensions,
which can be completely solved using their infinite
non-abelian symmetries, such as the Virasoro\ref\rbpz{\BPZ},
and the affine Lie algebra symmetries\ref\rkz{\KZ}.
In this paper we present a new application of affine Lie algebras to
massive 2d quantum field theory.

Though massive integrable quantum field theories have previously
been studied extensively, many of their important properties, such
as their correlation functions, have proven to be beyond computation
using the existing methods.  The exact method of quantum inverse
scattering\ref\rfst{\STF}\ provides an algebraic
foundation for the Bethe
ansatz.  There the emphasis is on the infinite number of local
integrals of motion $P_n$, which satisfy a trivial abelian algebra:
$[P_n , P_m ] = 0$.  This lack of algebraic structure in these conserved
quantities makes it impossible for example to obtain constraints on correlation
functions from Ward identities.  One is therefore led to the search
for interesting non-abelian conserved charges.  An integrable quantum
field theory with massive particles in its spectrum is characterized by
a factorizable S-matrix which must satisfy the Yang-Baxter
equation\ref\rzz{\ZZ}.
Since genuine conserved charges must commute with the S-matrix, the
interesting non-abelian symmetries can in principle be deduced on
the mass shell by inspection of the known  S-matrix.

The $q$-deformations of affine Lie algebras, i.e. quantum affine algebras,
were originally invented
to provide an algebraic characterization of known solutions of the
Yang-Baxter equation\ref\rdrin{\Drinfeld}\ref\rjim{\Jimbo}.
If the solution to the Yang-Baxter equation is taken to be a
physical S-matrix, then this characterization is precisely a symmetry
criterion.  In this way, the on-shell quantum affine
symmetries of the S-matrix can be understood as the minimal symmetry that
is strong enough to fix the S-matrix up to overall scalar factors.
The quantum affine symmetry thus replaces the original bootstrap
principles that led to the S-matrix, since solutions of the symmetry
equations automatically satisfy the Yang-Baxter equation.
Quantum affine algebraic structures
 also exist in the quantum inverse
scattering method; indeed this was the context in which they were originally
discovered. However we emphasize that the latter structures are not
symmetries in the conventional sense and that the quantum affine symmetry of
the
S-matrix has a completely different and independent physical content.
The quantum monodromy matrix has a smooth classical limit, whereas the
quantum affine charges do not.

Understanding the full implications of the quantum affine symmetry
requires an off-shell understanding of the
symmetry, namely, the conserved currents should be explicitly constructed.
Consider the sine-Gordon (SG) theory, which is the relevant example for
this paper.  The action is
\eqn\eIi{
S = \inv{4\pi} \int d^2 z \(  \d_z \phi \d_\zb \phi
                 + 4\la  \cos ( \betah \phi ) \) . }
In \ref\rbl{\BLnlc}\ref\rfl{G. Felder and A. LeClair,
Int. Journ. Mod. Phys. A7 Suppl. 1A (1992) 239.},
explicit conserved currents were constructed for the
$6$  generators corresponding to the simple roots of
the $q-\hat{sl(2)}$ affine Lie algebra, where
$q = \exp ( -2\pi i / \betah^2 )$.  In this realization
the central extension, or level,  is zero, and the symmetry is
actually a deformed loop algebra.  The construction of these
currents is purely quantum mechanical, and the resulting
charges do not have a smooth classical ($\betah \to 0$) limit.
The S-matrix for the
soliton scattering is the minimal solution to the quantum affine
symmetry equations.  This symmetry is somewhat exotic, in that the conserved
charges have in general fractional Lorentz spin $\pm ( 2/\betah^2 -1 )$.
Though the S-matrix was known in this case, this method was used to determine
the S-matrices in other models where it was not known, such as the affine Toda
theories, and perturbed minimal conformal models.

Using the quantum affine symmetry to compute properties beyond the S-matrix
in even the SG theory has proven to be difficult.  We now  believe this has
been
largely
due to having no  understanding of the role of affine Lie algebras in
the {\it massive} theory which occurs at $q=1$.
The primary aim of this paper is to properly develop
some new structures in the $q=1$ theory,
and more importantly, structures that can and will eventually be
$q$-deformed.
We emphasize that the deformation parameter $q$ has
nothing to do with perturbation away from the massless conformal limit.
Thus, though one can formally q-deform the structures of conformal field theory
by replacing affine Lie algebras with their $q$-analog, in doing this
the connection of these constructions with massive quantum field theory
is largely lost.

Using the quantum affine algebras, Frenkel and Reshetikhin formally
defined and
studied some $q$-analogs of conformal vacuum expectation
values\ref\rfr{I. B. Frenkel and N. Yu.
Reshetikhin, Commun. Math. Phys. 146 (1992) 1.}.  These satisfy
difference equations which can be viewed as
$q$-deformations of the Knizhnik-Zamolodchikov equations which
govern physical correlation functions of fields in conformal field
theory.  Smirnov used some aspects of this structure to reinterpret the basic
form-factor axioms from quantum affine symmetry\ref\rsmir{F. A.
Smirnov, Int. J. Mod. Phys.
A7, Suppl. (1992) .}\foot{The
symmetry considered in \rsmir\ is actually the Yangian symmetry, which occurs
at the $sl(2)$ invariant point of the SG theory at $\betah = \sqrt{2}$,
and is reached in
the limit  $q \to  -1$ of the quantum affine algebra. See \ref\ryang{D.
Bernard, Commun. Math. Phys. 137 (1991) 191.}\rbl.}.
(See also \ref\rdouble{D. Bernard and A. LeClair, {\it The Quantum Double
in Integrable Quantum Field Theory}, to appear in Nucl. Phys. B.}.)
However a direct link between
the results in \rfr\ and the physical form factors is missing.
One of the difficulties encountered is that
the Frenkel-Reshetikhin construction
generally utilizes
the infinite highest weight representations
of non-zero level $k$,
whereas the physical quantum affine symmetries studied thus far
in quantum field theory all have
zero level.

That the global quantum affine symmetry of the S-matrix necessarily has
zero level is easily understood.  The Hilbert space
$\hp$ of the theory has a multiparticle description, where states
diagonalize the momentum operators $P_\mu$.  At fixed particle number,
$\hp$ is a finite dimensional vector space depending on the
continuous momentum parameters, such as rapidity.  Given an algebra of
conserved charges that commute with the particle number operator, its
representation on $\hp$ must be the direct sum of finite dimensional
representations which are tensor products of the 1-particle representation.
Only the level zero affine algebras, or loop algebras,
have finite dimensional representations.
In a specific physical realization, the loop parameter is related to the
rapidity in a computable way.  Finally, since the quantum affine symmetry
charges commute with $P_\mu$, they can only relate states of the same
energy and thus cannot generate the spectrum of $\hp$.

It is  important to understand the significance of
non-zero level quantum affine algebras in massive integrable
field theory.  Having such algebras would entail the application of
the rich representation theory of these algebras.  Physical reasoning
provides an answer to this question.  Instead of $\hp$, consider the
space of fields.  The action of fields evaluated at the origin of space-time
on the vacuum $\Phi (0) \rvac$ defines a vector space $\hf$. The inner
products of states in $\hf$ with states in $\hp$ are the form-factors.
The space $\hf$ is a discrete space depending on no continuous parameters.
 Let $L$ be the generator of Lorentz boosts, or Euclidean rotations
in the space-time plane.  Each state in $\hf$ has a well-defined $L$
eigenvalue.  The operators $L, P_\mu$ comprise the Poincar\'e algebra:
$$[L, P_z ] = P_z , ~~~~~
[L, P_\zbar ] = - P_\zbar ,~~~~~  [P_z , P_\zbar ]= 0.$$
Obviously the spaces $\hp$ and $\hf$ are not simultaneously diagonalizable.
As we will see, the infinite dimensional representations of the non-zero
level affine Lie algebras characterize the space $\hf$.

In this paper we deal with the
$\qsl$ symmetry of the SG theory when $q=1$.  Studying this case serves
the purpose of disentangling the conceptual issues from the technical
complications which occur when $q\neq 1$.   Fortunately, the
point $q=1$ occurs at $\betah = 1$, which is just the free
Dirac fermion point of the SG theory\ref\rcol{\Colemani}.
Nevertheless, this is far from an empty exercise.  Though this
theory is trivial from the fermionic description, form-factors
and correlation functions of the fields $\exp ( i \al \phi )$
for $\al \notin \Zmath$
are not so trivial, since these fields are not simply expressed
in terms of the free fermions.
Surprisingly,
the $\slh$ symmetry of this free fermion
theory has not been considered before.

We now summarize the main results of this paper.
We first  construct the full infinite set of conserved $\slh$ charges
directly in the free massive
Dirac theory, and also the usual  infinite number of
abelian conserved charges $P_n$.  Radial quantization
is introduced as the natural way to obtain operators which
diagonalize $L$.  This introduces a fermionic fock space
description of $\hf$.  Furthermore, $\hf$ factorizes into
$\hf^L \ot \hf^R$, where in the massless limit\foot{In this paper
the massless limit always corresponds to the ultraviolet conformal field
theory, and `massless' and `conformal' will be used interchangeably.
},
$\hf^L$ ($\hf^R$) is the left (right) `moving' space of states.
Techniques are developed for studying this
`quasi-chiral factorization' in momentum space, including
the definition of an operator product expansion in this space.
This leads to a new and very simple way to derive the form factors
of the fields $\exp (\pm i \phi /2 )$ in the SG theory, as
momentum space correlation functions.
In section 5, we show how the conserved charges also factorize in
their action on $\hf^L \ot \hf^R$.  This leads to two separate
algebras $\hat{sl(2)}_L$ and $\hat{sl(2)}_R$, which each have level $1$.  In
the
same fashion, the integrals of motion $P_n^{L,R}$ satisfy an infinite
Heisenberg algebra.  It is shown how the spectrum of $\hf$
can be obtained by supplementing the $\slh$ algebra with this
Heisenberg algebra, the fields being organized
into infinite highest weight modules.
We use the integrals of motion to formulate an exact momentum space
bosonization.  In this operator formulation, non-trivial SG form-factors
are computed as expectation values of vertex operators between
level 1 highest weight states.

\bigskip
\newsec{Affine $\hat{sl(2)}$ Symmetry of the Massive Dirac Fermion}

The Dirac theory is a massive free field theory of charged fermions.
Introducing the Dirac spinors
$ \Psi_\pm = \left(\matrix{\psib_\pm \cr \psi_\pm \cr}\right)$
of $U(1)$ charge $\pm 1$, with the appropriate choice of
$\gamma$-matrices\foot{
$\gamma^0 = \left(\matrix{ 0 & -i \cr i & 0\cr}\right) ,
{}~~\gamma^1 = \left(\matrix{ 0 & i \cr i & 0 \cr}\right) .$}
the action reads
\eqn\eIIi{
S = - \inv{4\pi} \int dx dt \(
\psibm \d_z \psibp + \psim \d_\zb \psip
+ i \ma ( \psim \psibp - \psibm \psip ) \) . }
We have continued to Euclidean space $t \to -it$, and $z,\zb$ are the
usual Euclidean light-cone coordinates:
\eqn\eIIii{
z = \frac{t + ix}{2} , ~~~~~\zb = \frac{t-ix}{2} . }
Canonical quantization gives
\eqn\eIIiii{
\left\{ \psip(x,t), \psim (y,t) \right\} = \left\{
\psibp (x,t) , \psibm (y,t) \right\}
= 4\pi \de (x-y) , ~~~\left\{ \psi (x,t) , \psib (y,t) \right\} = 0. }

Generally, the conserved quantities follow from conserved currents
$J_\mu$:
\eqn\eIIiv{
\d_\zb J_z  + \d_z J_\zb = 0. }
The usual time-independent conserved charges are then
\eqn\eIIivb{
Q= \inv{4\pi} \int dx \> \( J_z (x) + J_{\zbar} (x) \) . }
In the sequel, we will often display    the two components
of conserved currents by writing
\eqn\ecurrent{
Q = \int \dz ~  J_z ~ - ~ \int \dzb ~ J_\zb , }
i.e. without specifying the contour of integration.
The standard conserved charges in the Dirac theory are the $U(1)$ charge $T$,
\eqn\eIIv{
T = \int \dz ~\psip \psim ~ - ~ \int \dzb ~ \psibp \psibm , }
and the Poincar\'e generators $L, P_z , P_\zb$.  The operator $L$ generates
Euclidean rotations (Lorentz boosts), and $P_z , P_\zb$ generate
space-time translations, where the Hamiltonian is $P_z + P_\zb$.
On fields, $P_z = \d_z , ~ P_\zb = \d_\zb$.

The Dirac theory has an infinite number of additional conserved quantities,
which we now construct.  Though, as we will see, these can be constructed
directly in the Dirac theory, it is interesting to see how some of them arise
from the $q\to 1$ limit of the results in \rbl, as the eventual goal is
to go beyond $q=1$.   There, four non-local charges were constructed in the
SG theory for any value of the coupling $\betah$.  They were constructed
using the methods of conformal perturbation theory developed generally
by Zamolodchikov\ref\rcpt{\Zamoiii}.  Let $\vphi^L, \vphi^R$
denote the quasi-chiral components of the SG field $\phi$, such that in
the massless ultraviolet limit
\eqn\eIIvi{
\phi ~ \mtoo ~ \vphi^L (z) + \vphi^R (\zb ) . }
The exact propagator in this limit is
\eqn\eprop{
\lvac \phi (z,\zb ) \phi (0) \rvac = - \log (z \zb /R^2 ) , ~~~~~~(\ma = 0 ) }
where $R$ is an infrared cutoff.
At the free fermion point $\betah = 1$, the non-local charges of \rbl\ take the
following form:
\eqn\eIIviii{
\eqalign{
Q_\pm &= \int \dzf ~\exp\( {\pm 2 i \vphi^L }\) - \lambda \int \dzbf
{}~ \exp ( {\pm i \vphi^L \mp i \vphi^R} ) \cr
\bar{Q}_\pm &= \int \dzf \lambda  ~ \exp ( {\mp i \vphi^R \pm i \vphi^L }) -
\int \dzbf
{}~ \exp ({\mp 2 i \vphi^R})  .\cr }}
All field  operators are well-defined in the framework of conformal
perturbation
theory\foot{
Based on the work of Mandelstam\ref\rmand{\Mand},
in \rbl\ the following definition
of the quasi-chiral components was proposed:
\eqn\eIIvii{\eqalign{
\vphi^L (x,t) &= \inv{2} \( \phi (x,t) + \int_{-\infty}^x dy \>
\d_t \phi (y,t) \)
\cr
\vphi^R (x,t) &= \inv{2} \( \phi (x,t) - \int_{-\infty}^x dy \>
\d_t \phi (y,t) \)
.\cr }}
The conservation of the above non-local charges in a canonical equal-time
framework was studied in \ref\rrajab{\Rajab}\ref\rFF{\FF}, where it was found
that the currents in this framework differ slightly from the
expression obtained by literally substituting the expressions
\eIIvii\ into \eIIviii.}, and are implicitly normal ordered.

It was noted in \ref\rraja{\Raja}\ that at the special values of the
SG coupling constant $\betah = 1/n$, the non-local currents can be expressed
in terms of the Thirring fermions.  At $\betah = 1$ the non-perturbative
identification of the terms in the action is\rcol
\eqn\boson{
\la ~ : \cos (\phi ) : = -i \ma ~ ( \psim \psibp - \psibm \psip ) .}
It is a simple matter to use the
bosonized relations\rmand
\eqn\eIIix{
\psi_\pm  = \exp ( {\pm i \vphi^L} ), ~~~~~\psib_\pm = \exp ({\mp i \vphi^R })
}
in the expressions \eIIviii\ to yield:
\eqn\eIIx{
\eqalign{
Q_\pm &= \int \dzf (\psi_\pm \d_z \psi_\pm )  -  \int \dzbf
(i\ma \psib_\pm \psi_\pm )  \cr
\bar{Q}_\pm &= \int \dzf (-i \ma \psi_\pm \psib_\pm )  -
\int \dzbf
(\psib_\pm \d_\zb \psib_\pm )  .\cr }}

In this fermionic description, one sees that the currents are local,
and that they have the structure of the energy-momentum tensor, except
that they carry $U(1)$ charge $\pm 2$.
That the currents for the charges $Q_\pm , \bar{Q}_\pm $
are conserved (eq. \eIIiv) is now a simple consequence of the
equations of motion:
\eqn\eIIxi{
\d_z \psib_\pm = i \ma \psi_\pm ,~~~~~
\d_\zb \psi_\pm = -i \ma \psib_\pm . }
It is interesting to note that whereas the conservation of the
charges \eIIviii\ in the bosonic SG description is a purely
quantum mechanical phenomenon (verification of their conservation
by construction uses the operator product expansion, and furthermore,
in the classical limit $\betah \to 0$ the charges are ill-defined due
to the $1/\betah$'s appearing in the exponentials), in the fermionic
description the fermionic statistics introduces just enough
quantum mechanics so that the conservation of the charges is now purely
classical in origin.

Having understood the classical origin of the conservation of the above
charges, it is now straightforward to find an infinite number of
additional quantities which are also conserved as  a simple consequence of
the equations of motion \eIIxi.  They are the following:
\eqn\eIIxii{\eqalign{
\Qt^\pm_{-n} &= \frac{(-1)^{n+1}}{2}
\( \int \dz ( \psi_\pm \d_z^n \psipm ) - \int \dzb ( i\ma \> \psibpm \d_z^{n-1}
\psipm ) \)
\cr
\Qt^\pm_n &= \frac{(-1)^{n+1}}{2}
\( \int \dz ( -i\ma \> \psipm \d_\zb^{n-1} \psibpm )
- \int \dzb ( \psibpm \d_\zb^n \psibpm ) \)
\cr
\al_{-n} &= (-)^n \(
\int  \dz ( \psip \d_z^n \psim ) - \int \dzb ( i\ma \>\psibp \d_z^{n-1} \psim )
\)
\cr
\al_n &= (-)^n \(
\int \dz ( -i\ma \>  \psip \d_\zb^{n-1} \psibm )
- \int \dzb ( \psibp \d_\zb^n \psibm ) \)
, \cr } }
where $n\geq 1$ is an integer.

The previous charges are identified as $Q_\pm = \Qt^\pm_{-1}$ and
$\bar{Q}_\pm = \Qt^\pm_1$.  The ordinary momentum operators are
identified as
\eqn\eIIxiii{P_z = \al_{-1},~~~~~P_\zb = \al_1 . }
For convenience of notation we define
\eqn\eIIxiv{
\al_0 \equiv T.}
 From the hermiticity properties $(\psipm)^\dag = \psi_\mp$,
$(\psib_\pm )^\dag = \psib_\mp$, one finds
\eqn\eIIxv{
\( \Qt^\pm_n \)^\dag = \Qt^\mp_n , ~~~~~\al_n^\dag = \al_n .}

In order to simplify the computation of the algebra of these
charges, we translate them to on-shell momentum space.
Introducing a rapidity variable $\th$ which parameterizes on-shell
momentum,
\eqn\eIIxvi{
p_z (\th ) = \ma e^\th , ~~~~~p_\zb (\th ) = \ma e^{-\th} , }
the conventional expansions are the following:
\eqn\eIIxvib{\eqalign{
\psip (x,t) &= i \sqm \int_{-\infty}^\infty
\frac{d\th}{2\pi i} ~ e^{\th/2}
\( c(\th ) \> e^{-ip(\th ) \cdot x }  - d^\dag (\th )
\> e^{i p (\th)  \cdot x } \)
\cr
\psibp (x,t) &=  \sqm \int_{-\infty}^\infty
\frac{d\th}{2\pi i} ~ e^{-\th/2}
\( c(\th ) \> e^{-ip(\th ) \cdot x }  + d^\dag (\th ) \> e^{i p(\th )
\cdot x } \)
,\cr}}
with $\psim = \psip^\dag , \psib_- = \psibp^\dag$.   The momentum-space
operators satisfy
\eqn\eIIxvii{
\{ d(\th), d^\dag (\th' ) \} = \{ c(\th ) , c^\dag (\th' ) \}
= 4 \pi^2 \de (\th - \th' ) . }

To further simplify the computation, and also in anticipation of the
following section, we combine creation and annihilation operators into
a single operator as follows.  Define the momentum space variable $u$
as
\eqn\eIIxviib{
u\equiv e^\th , }
and let
\eqn\eIIxviii{\eqalign{
\bhp (u) &= d^\dag (u)  ,
{}~~~~~~~~~~\bhm (u) = c^\dag (u)  ~~~~~~~~~~~{\rm for} ~ u>0
\cr
\bhp (u) &= i c(-u)   ,
{}~~~~~~~~\bhm (u) = i d(-u)    ~~~~~~~~~{\rm for} ~ u<0
.\cr }}
These operators satisfy
\eqn\eIIxix{
\eqalign{
\{ \bhp (u) , \bhm (u') \} &= 4 \pi^2 i |u| \de (u + u') , ~~~~~~
\{ \bh^\pm (u) , \bh^\pm (u') \} = 0 \cr
\( \bhp (u) \)^\dag &= -i \bhm (-u) . \cr }}
Then,
\eqn\eIIxx{
\Psi_\pm = \col{\psibpm}{\psipm} = \pm \sqm
\int_{-\infty}^\infty \du \> \bh^\pm (u) ~
\col{1/ \squ}{-i\squ} ~ \ez . }

The charges now have the simple momentum space expressions:
\eqn\eIIxxi{\eqalign{
\Qt^\pm_n &= \frac{\ma^{|n|}}{4\pi} \int_{-\infty}^{\infty} \du
\> u^{-n} ~ \bh^\pm (u) \bh^\pm (-u) \cr
\al_n  &= \frac{\ma^{|n|}}{2\pi} \int_{-\infty}^{\infty} \du
\> u^{-n} ~ \bh^+ (u) \bh^- (-u) ,  \cr}}
for all $n$.
These expressions were derived by first expressing the charges \eIIxii\
in the form \eIIivb, and then substituting the expressions \eIIxx.
It is now straightforward to compute the
commutation relations of these charges using \eIIxix.  One finds
that the structure of the resulting relations depends significantly
on whether $n$ is odd or even.  First note that by making the
redefinition $u\to -u$ in \eIIxxi, and using \eIIxix, one finds
$$\Qt^\pm_n = (-1)^{n+1} \Qt^\pm_n ~~~\Rightarrow ~~~ \Qt^\pm_m
= 0 ~~~~{\rm for} ~ m ~ {\rm even} . $$
Define
\eqn\eIIxxii{\eqalign{
P_n \equiv \al_n , ~~~~~~Q^\pm_n \equiv \Qt^\pm_n , ~~~~~ &n ~ {\rm odd}
\cr
T_n \equiv \al_n , ~~~~~~~~~~~~ &n ~ {\rm even} . \cr}}
Then one finds the following algebraic relations
\eqna\eIIxxiii
$$\eqalignno{
\[ P_n , P_m \] &= 0  &\eIIxxiii {a} \cr
\[ P_n , T_m \] &= \[ P_n , Q^\pm_m \] = 0 &\eIIxxiii {b} \cr
\[ T_n , T_m \] &= 0 &\eIIxxiii {c} \cr
\[ T_n , Q^\pm_m \] &= \pm 2 ~ \mn ~ Q^\pm_{n+m}  &\eIIxxiii {d} \cr
\[ Q^+_n , Q^-_m \] &= \mn ~ T_{n+m} &\eIIxxiii {e} . \cr } $$

It will be important to introduce the conserved Lorentz boost operator
$L$;  in momentum space one finds
\eqn\eIIxxiv{
L = \inv {2\pi} \int_{-\infty}^\infty \du \>
\bh^+ (u) \> u\d_u \bh^- (-u) . }
The conserved charges have integer Lorentz spin:
\eqn\eIIxxv{
\[ L , \al_n \] = -n \> \al_n , ~~~~~\[ L , Q^\pm_n \] = -n \> Q^\pm_n . }

We now interpret this algebraic structure. The $P_n$'s are the usual infinity
of commuting integrals of motion with Lorentz spin equal to an odd integer,
where $P_z = P_{-1}, ~ P_\zb = P_1 $, and the hamiltonian is $P_1 + P_{-1}$.
These were already known to exist at all values of the SG coupling constant.
The additional charges $T_n , Q^\pm_n $
all commute with the $P_m$'s; for $m= \pm 1$ this is just the statement
that they are all conserved.  In the SG theory, the $T_n$'s are
generalizations of the topological charge $T_0$.

The commutation
relations of the $T_n, \Qpm n$ are the defining relations of the level
$0$ $\slh$ affine Lie algebra.  Since this realization of affine
Lie algebras is relatively unfamiliar in the physics literature in
comparison with their realization in current algebra, we clarify this
point\foot{For reviews of affine Lie algebras in mathematics and
physics see \ref\rkac{\Kac}\ref\rgo{\GO}.}.
The most economical definition of the affine Lie algebras involves only
the finite number of generators for the simple roots, $e_i , ~ f_i, ~ h_i$,
satisfying
\eqn\eIIxxvi{\eqalign{
[h_i , e_j ] &= a_{ij} e_j , ~~~~~[h_i , f_j ] = - a_{ij} f_j ~~~~~
[e_i , f_j ] = \de_{ij} h_i \cr
ad^{1-a_{ij} } e_i (e_j ) &
= ad^{1-a_{ij}} f_i (f_j ) = 0 ~~~~~~(i\neq j )
,\cr } }
where $a_{ij}$ is the generalized Cartan matrix.  For $\slh$, $i,j \in \{ 0,1
\} $, and $a = \mat{2}{-2}{-2}{2}$.

The algebra \eIIxxvi\ defines an infinite dimensional Lie algebra.
Though conformal current algebra is not relevant for the problem we
are considering, it is nevertheless useful to recall how the infinite
algebra arises in that context.
There one deals with an $sl(2)_L \ot sl(2)_R$ invariant
theory, with chiral currents $j^a (z), ~\bar{j}^a (\zb )$.  Upon making the
analytic conformal transformation $z \to \log (z)$, the currents can be
expanded in modes $j^a (z) = \sum_n j^a_n z^{-n-1}$, and similarly for
$\bar{j}^a_n$.  Introducing the derivation $d$, these modes satisfy
\eqn\eIIxxvii{\eqalign{
[ j^0_n , j^\pm_m ] &= \pm 2 \>  j^\pm_{n+m} , ~~~~~[j^0_n , j^0_m ] =
2kn \>
\de_{n, -m}  \cr
[j^+_n , j^-_m ] &= j^0_{n+m} + kn \> \de_{n, -m} \cr
[d , j^a_n ] &= n \> j^a_n , \cr }}
where the central extension $k$ is called the level.
The algebras defined by \eIIxxvi\ and \eIIxxvii\ are identical.
The simple root generators are given by
\eqn\eIIxxviib{
\eqalign{
e_1 &= j^+_0 , ~~~~ f_1 = j^-_0 , ~~~~h_1 = j^0_0 \cr
e_0 &= j^-_1 , ~~~~f_0 = j^+_{-1} , ~~~~ h_0 = -j^0_0 + k . \cr }}

The algebra \eIIxxvii\ cannot describe a symmetry algebra for the
Dirac theory since it has an $sl(2)$ subalgebra generated by the
Lorentz scalars $j^a_0$, whereas the Dirac theory has only a $\uone$
global symmetry.  What is relevant for the Dirac theory is actually
a twisted $\slh$ affine algebra.  Consider the inner automorphism $\tau$
of $sl(2)$, $\tau (j^a_0 ) = e^{i\pi j^0_0 / 2 } \( j^a_0 \)
e^{-i\pi j^0_0 /2 } $.  This $\tau$ can be used to construct an inner
automorphism of the algebra $\slh$ by the formula
\eqn\eIIxxviii{
t^a (v ) = \sum_m t^a_m v^m =
v^{j^0_0 /2} j^a (v^2 ) v^{-j^0_0 /2 } . }
Thus, defining
\eqn\eIIxxix{\eqalign{
t^\pm_n &= j^\pm_{(n\mp 1)/2} , ~~~~~~~~~ n ~ {\rm odd} \cr
t^0_n &= j^0_{n/2} -  \frac{k}{2}
\> \de_{n,0}  ~~~~~ n ~ {\rm even} \cr
d' &= 2d + j^0_0 /2 ,
\cr }}
one finds
\eqn\eIIxxx{\eqalign{
[t^0_n , t^0_m ] &= k \> n \> \denm , ~~~~~[t^0_n , t^\pm_m ] = \pm 2
\> t^\pm_{n+m} \cr
[t^+_n , t^-_m ] &= t^0_{n+m} +  \frac{kn}{2} \>
\denm  \cr
[ d' , t^a_n ] &= n \> t^a_n . \cr }}
This twisted $\slh$ algebra is of course isomorphic to \eIIxxvii\ by
construction.  The simple root generators are now
\eqn\eIIxxxb{
\eqalign{
e_1 &= t^+_1  , ~~~~ f_1 = t^-_{-1} , ~~~~h_1 = t^0_0 + \frac{k}{2}
\cr
e_0 &= t^-_1 , ~~~~f_0 = t^+_{-1}  , ~~~~ h_0 = -t^0_0 + \frac{k}{2}
. \cr }}
In the mathematics literature, the algebra \eIIxxvii\ is referred to as
being in the homogeneous gradation, whereas \eIIxxx\ is in the
principal gradation.  Note that in going from the homogeneous to the
principal gradation the $sl(2)$ global zero mode algebra is broken to
$\uone$ generated by $t^0_0$.

The algebra of the charges $T_n , ~ \Qpm n$ is the same as for the twisted
$\slh$ algebra at level $k=0$ if one absorbs the mass dependence into
the definition of the charges: $t^\pm_n = \Qpm n \> \ma^{-|n|} ,
t^0_n = T_n \> \ma^{-|n|}$, or simply by a choice of units sets $\ma = 1$.
One also has the identification $d' = L$.  This implies that in the
massless limit $\ma \to 0$ one doesn't recover an $\slh$ algebra.
Rather, in the massless limit one obtains two decoupled
Borel subalgebras generated by $\{ e_i ,  h_i \}$ and
$\{ f_i ,  h_i \} $ respectively.
  Note further that in position space the
charges \eIIxii\ do not correspond to moments of a universal operator as
in conformal field theory, however in momentum space they do.

The conserved charges have the following simple commutation
relations with the operators $\bh^\pm (u)$:
\eqn\eIIxxxi{\eqalign{
\[ \al_n , \bh^\pm (u) \] = (\pm )^{n+1} \ma^{|n|} u^{-n} \> \bh^\pm (u)
{}~~&~~~\[ Q^\pm_n , \bh^\mp (u) \] = \ma^{|n|} u^{-n} \> \bh^\pm (u) \cr
\[ \Qpm n , \bh^\pm (u) \] &= 0 . \cr }}
This implies that on the doublet of one particle states
$\col{ d^\dag (\th ) \rvac}{c^\dag (\th ) \rvac}$, the charges have the
following loop algebra representation
\eqn\eIIxxxii{
\al_n = \ma^{|n|} u^{-n} \mat{1}{0}{0}{ (-1)^{n+1} } , ~~~
Q^+_n = \ma^{|n|} u^{-n} \mat{0}{1}{0}{0} , ~~~
Q^-_n = \ma^{|n|} u^{-n} \mat{0}{0}{1}{0} . }

 From the equations \eIIxxxi\ and \eIIxx\ one finds the following action
on the fermion fields:
\eqn\eIIxxxiii{\eqalign{
[\al_n , \psipm ] &= -i \ma (\pm )^{n+1} \d_\zb^{n-1} \psibpm ,
{}~~~~~[\al_{-n} , \psipm ] = (\pm )^{n+1} \d_z^n \psipm \cr
[\al_n , \psibpm ] &= (\pm )^{n+1} \d_\zb^{n} \psibpm ,
{}~~~~~~~~~~~[\al_{-n} , \psibpm ] = i\ma (\pm )^{n+1} \d_z^{n-1} \psipm \cr
[\Qpm n , \psi_\mp ] &= i\ma \> \d_\zb^{n-1} \psib_\pm , ~~~~~
{}~~~~~~~~[Q^\pm_{-n} , \psi_\mp ] = - \d_z^n \psipm  \cr
[\Qpm n , \psib_\mp ] &= - \d_\zb^n \psibpm , ~~~~~
{}~~~~~~~~~~~[Q^\pm_{-n} , \psib_\mp ] = -i \ma \> \d_z^{n-1} \psipm , \cr }}
for $n \geq 0$;  all other commutators are zero.

The algebra \eIIxxiii{}, along with the above action on the fermion
fields is sufficient to reconstruct the Dirac theory.  Introducing
the notation $Q(\psi ) = [Q, \psi ]$, one has
\eqn\eIIxxxiv{
Q_1^+ \( Q^-_{-1} \( \psi_+ \) \) = - \d_z Q^+_1 (\psi_- ) = -i\ma \d_z
\psibp . }
On the other hand, since $Q^+_1 (\psi_+ ) = 0$,
\eqn\eIIxxxv{
Q^+_1 \(  Q^-_{-1} \( \psi_+ \) \)  =
[Q^+_1 , Q^-_{-1} ] \( \psip \) = \ma^2 T_0 (\psip ) = \ma^2 \psip . }
Comparing the last two equations, one finds the Dirac equation
$\d_z \psibp = i\ma \psip$.

We now illustrate how Ward identities for the $\slh$ algebra fix the
2-point correlation functions of the fermions.  Certainly, these correlation
functions are known from far less sophisticated reasoning.  We go through
this exercise merely to confirm that it is indeed possible.
All of the charges annihilate the vacuum, thus the Ward identities take
the form
\eqn\eIIxxxvi{
\lvac Q \( \psi_1 (x_1 ) \) \> \psi_2 (x_2 ) \rvac
+ \lvac \psi_1 (x_1 ) \> Q \( \psi_2 (x_2 ) \) \rvac = 0 }
for any conserved charge $Q$.  One has
\eqn\eIIxxxvii{
\lvac Q^-_{-1} \( \psip (z, \zb ) \) Q^+_1 \( \psibm (0) \) \rvac
= - \d_z \d_\zb \> \lvac \psim (z, \zb ) \psib_+ (0) \rvac. }
On the other hand, from the Ward identity one has that the LHS of the
above equation equals
\eqn\eIIxxxviii{
-\lvac \psip (z, \zb ) Q^-_{-1} Q^+_1 \( \psibm (0) \) \rvac
= - \ma^2 \lvac \psip (z, \zb ) \psibm (0) \rvac . }
Comparing \eIIxxxvii\ and \eIIxxxviii, and using charge conjugation
symmetry, one finds
\eqn\eIIxxxix{
\( \d_z \d_\zb - \ma^2 \)
\lvac \psibm (z ,\zb ) \psip (0) \rvac = 0. }
Given the solution to
\eIIxxxix, the Dirac equation can then be used to find the other
2-point functions.  The result is
\eqn\eIIxxxx{\eqalign{
\lvac \psibm (z, \zb ) \psip (0) \rvac &= -2 i \ma K_0 (\ma r) \cr
\lvac \psim (z, \zb ) \psip (0) \rvac &= 2  \ma
\sqrt{\frac{\zb}{z} } K_1 (\ma r) \cr
\lvac \psibm (z, \zb ) \psibp (0) \rvac &= 2 \ma
\sqrt{\frac{z}{\zb} } K_1 (\ma r) , \cr }}
where $K_n$ is the standard modified Bessel function.

\bigskip
\newsec{Quasi-Chiral Factorization of the Space of Fields}

\bigskip
\noindent   3.1  {\it ~~General Remarks}
\medskip

The origin of the quasi-chiral factorization we will introduce
in the Dirac theory can be understood in a more general setting.
In conventional approaches to free or interacting massive
quantum field theory one constructs an asymptotic free particle
fock space $\CH_P$.  These states diagonalize the momentum
operators $P_z , ~ P_\zb$, whose eigenvalues are parameterized by the
rapidity of the particles. On 1-particle states $P_z = \ma \, u,
P_\zb = \ma / u$.  To study form-factors and correlation functions
one considers the space of fields $\CH_F$:
\eqn\eIIIi{
\hf \equiv \left\{ \vert \Phi \rangle \equiv \Phi (z, \zb = 0 ) \rvac \right\}
, }
where $\Phi (z , \zb )$ is any field in the theory.  The spaces
$\hp$ and $\hf$ are completely different;  the inner product of
states in $\hp$ with states in $\hf$ are the form-factors, which
are generally non-trivial\foot{See \ref\rform{\form}\ for the construction
of the multiparticle form-factors in a variety of models, including
the SG theory.}.

Since $P_{z , \zb}$ act as spacetime derivatives on fields, the states in
$\hf$ obviously do not diagonalize $P_{z, \zb}$.  Rather, since
every field has well-defined properties under Euclidean rotations,
the space $\hf$ diagonalizes the Lorentz boost operator $L$:
\eqn\eIIIii{
L \vert \Phi \rangle = s~ \vert \Phi \rangle , }
where $s$ is the Lorentz spin of the field $\Phi$.

In relativistic quantum field theory, the standard equal-time, planar
quantization (e.g. \eIIiii) yields the space $\hp$.  To construct the
space $\hf$, one should consider instead radial quantization.  Define
the radial coordinates $(r, \vphi )$ as follows:
\eqn\eIIIiii{
z = \frac{r}{2} ~ \e{i\vphi} , ~~~~~\zb = \frac{r}{2} ~ \e{-i \vphi } . }
In radial quantization one treats the $r$-coordinate as a `time', and
$\vphi$ as the `space'.  Equal-time commutation relations are specified
along circles surrounding the origin.  For example, in the Dirac theory,
rewriting the action in terms of the $r, \vphi$ coordinates using
\eqn\eIIIiv{
\d_z = e^{-i\vphi} ( \d_r -  {\frac{i}{r}} \d_\vphi )
, ~~~~~\d_\zb  = e^{i\vphi} ( \d_r +  {\frac{i}{r}} \d_\vphi ), }
and applying canonical quantization yields
\eqn\eIIIv{
\{ \psip (r, \vphi ) , \psim (r , \vphi' ) \} =
\frac{4\pi}{r} \e{i\vphi} \de ( \vphi - \vphi' ) ,
{}~~~~~
\{ \psibp (r, \vphi ) , \psibm (r , \vphi' ) \} =
\frac{4\pi}{r} \e{-i\vphi} \de ( \vphi - \vphi' ) . }

As a differential operator
on the spacetime coordinates, $L= -i \d_\vphi$,
thus the states constructed in radial quantization are eigenstates of
$L$.  The Poincar\'e algebra does not provide any additional
quantum numbers.  In order to find additional quantum numbers of the space
$\hf$, we consider the following.
In the usual planar quantization, the states in $\hp$ are created
by asymptotic `in' or `out' fields, in the sense of scattering theory.
The mass-shell condition arises from the Klein-Gordon equation satisfied
by these fields: $( \d_z \d_\zb - \ma^2 ) \Phi_{{\rm in , out}} = 0$.
Consider now the analog of these asymptotic
fields in radial quantization.  Defining the scaling operator
$D= r\d_r$, the Klein-Gordon equation reads
\eqn\eIIIvi{
(D+L)(D-L) \Phi (r, \vphi ) = \ma^2 r^2 ~ \Phi (r, \vphi ) . }
This implies
\eqn\eIIIvii{
(D+L) (D-L) ~ \Phi (r=0 ) = 0. }
Assuming the space $\hf$ can be constructed from the
analog of these asymptotic fields, one has the factorization
\eqn\eIIIviii{
\hf = \hf^L \ot \hf^R , }
where
\eqn\eIIIix{
(D-L) \vert \hf^L \rangle = (D+L) \vert \hf^R \rangle = 0. }
Thus, the quasi-chiral factorization arises as the radial analog
of the mass shell condition.
In massive quantum field theory, $D$ is not a quantum conserved
operator; nevertheless, every field $\Phi$ has a well-defined
scaling dimension $\De_\Phi$, and we define $D\vert \Phi \rangle
= \De_\Phi \> \vert \Phi \rangle $.

The manner in which the mass $\ma$ enters the equation \eIIIvi\
indicates that the space $\hf$ can also be obtained by taking the
$\ma \to 0$ limit.  More precisely, let $\hf^{(0)}$ denote the
space $\hf$ modulo identifications resulting from the
mass-dependent equations of motion.  For example, in the
Dirac theory, since $\d_z \psibp = i\ma \psip$, the two fields
$\d_z \psibp$ and $\psip$ are linearly related, and $\psip$ is
defined to be
in $\hf^{(0)}$ but $\d_z \psibp$ is not.  I.e. we define
$\hf^{(0)}$ to be a linearly independent basis for the space
$\hf$, where each field in $\hf^{(0)}$ has no explicit mass dependence.
Then for $\Phi (0) \rvac \in \hf^{(0)}$,
\eqn\eIIIx{
\lim_{r\to 0} ~~ \Phi (r) \rvac = \lim_{r\to 0} ~ \lim_{\ma \to 0}
\Phi (r) \rvac . }

The significance of the equation \eIIIx\ is that it implies the
structure of $\hf^{(0)}$ is identical to that in the $\ma \to 0$
limit.  In this limit $\hf^L$ and $\hf^R$ are the well-understood
left and right `moving' space of states in conformal field
theory\rbpz.  We emphasize however that the fields in
$\hf^L$ $(\hf^R )$ are obviously not solely functions of
$z$ $(\zb )$ in the massive theory.

In the above discussion, there are some hidden assumptions about
the renormalization group properties of the theory that we now
clarify.  The massive theories we have in mind can generally be formulated
as perturbed conformal field theories \rcpt, with the formal action
\eqn\eIIIxi{
S = S_{CFT} + \sum_i \frac{\la^i}{2\pi}  ~ \int d^2 z ~ \CO_i (z , \zb ),}
where $\CO_i$ are relevant operators with scaling dimension $\leq 2$.
We assume there is no wavefunction renormalization, so that the anomalous
scaling dimension of  a field is constant along the renormalization
group trajectory.  In the SG theory for example, this situation occurs
since the theory may be renormalized by simply normal ordering the
$\cos (\betah \phi )$ potential and redefining $\la$\rcol.
This means that the theory cannot have any non-trivial infra-red
fixed points.  Since the scaling dimension of a field is independent
of the $\la^i$ under these assumptions, the scaling dimensions of
fields and coupling constants $\la^i$ is fixed by the properties
of the ultraviolet conformal field theory.  The physical mass scale
$\ma$ is a function of the $\la^i$, and physical correlation
functions are expressed in terms of $\ma$.  In summary, we are making
the same basic assumptions as in \rcpt, i.e. that the space of fields
in the massive theory has the same structure as in the conformal field
theory.  Radial quantization provides a way of constructing the space
$\hf$ with these structures explicitly, with the $L$ and $D$ quantum numbers
the same as in the conformal limit\foot{
In the conformal limit, the usual Virasoro zero modes are
$$L_0 = \frac{D+L}{2} , ~~~~~\bar{L}_0 = \frac{D-L}{2} . $$ }.

It will be important for us to understand the above statements in momentum
space.  As we will show by explicit construction,
radial quantization leads to momentum
space operators $\hat{\Phi}^L (u)$ and $\hat{\Phi}^R (u)$ with
analytic expansions in the variable $u$.  Since $P_z = \ma u,
{}~ P_\zb = \ma /u $, the limit $u\to \infty$ sets $P_\zb = 0$, whereas
$u\to 0$ sets $P_z = 0$.  Thus the spaces $\hf^{L,R}$ can be constructed
in momentum space as follows:
\eqn\eIIIxii{
\hf^L = \left\{ \hat{\Phi}^L (u \to \infty ) \rvac \right\}
, ~~~~~\hf^R = \left\{ \hat{\Phi}^R (u\to 0 ) \rvac \right\} . }
Other general aspects of this quasi-chiral structure will be developed
by example.

\bigskip
\noindent 3.2 {\it ~~Free Dirac Fermions}
\medskip

We now apply the above ideas to the free Dirac fermion theory.  Some
aspects of radial quantization in this situation were developed
in \ref\rti{\TI}\ref\rgrif{\Griffin}.

The simplest way to formulate radial quantization is through
analytic properties in the momentum space $u$-variable.  The
$\bh^\pm (u)$ expansion for the fermion fields \eIIxx\ is constructed
to satisfy the equations of motion.  In radial quantization, the
equations of motion are unchanged, only the canonical quantization
procedure differs.  This means that in radial quantization the expressions
\eIIxx\ are still valid with the appropriate modification of the
integral over $u$.  In radial quantization the angular variable
$\vphi$ is treated as a space variable, thus the $u$-integral should be
analytically continued to a circle surrounding the origin in the
complex $u$ plane.  Let us now suppose that $\bh^\pm (u)$ is single valued
in the $u$ plane.  The branch cuts due to $\squ$ in \eIIxx\ then imply
that the fermion fields are anti-periodic:
$\psi (r, \vphi + 2\pi i ) = - \psi (r, \vphi )$.
There are precisely two ways of avoiding the cut to close the $u$ contour,
either toward $u=0$ or $u = \infty$.  We define the following prescription
for analytic continuation of the $u$ integral in \eIIxx:
\eqn\eIIIxiii{
\int_{-\infty}^\infty \du ~ \bhpm (u)
\rightarrow
\(
\int_{\cL} \dua ~ b_\pm (u) ~~ - ~~
\int_{\cR} \dua ~ \bb_\pm (u) \), }
where
$\cL , \cR$ are contours depending on the angular direction $\vphi$ of
the cut displayed in figures 1,2,  and $b_\pm (u)$ and
$\bb_\pm (u)$ are distinct operators defined explicitly below.
\midinsert
\epsfxsize = 2in
\vbox{\vskip -.1in\hbox{\centerline{\epsffile{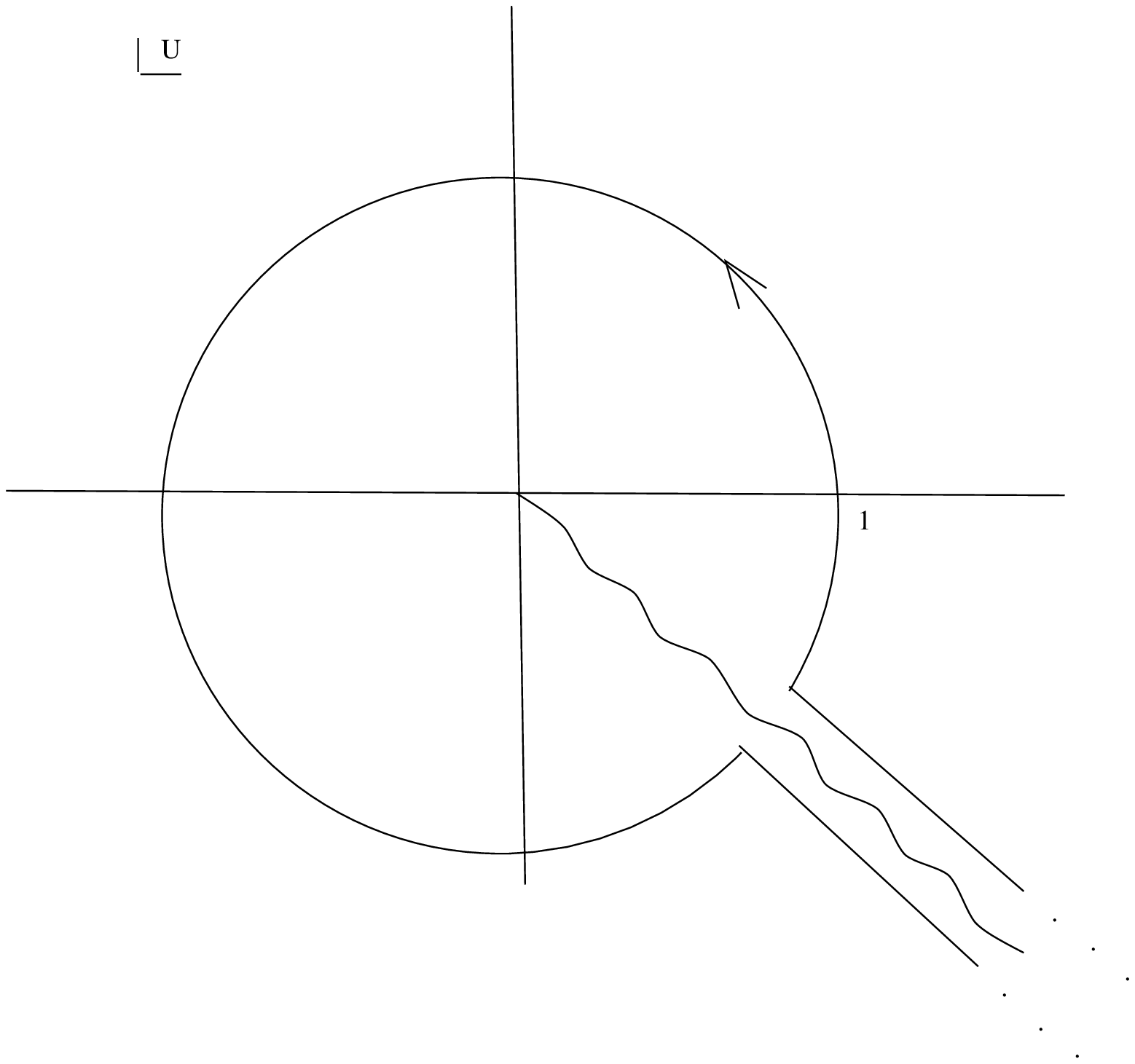}}}
\vskip .1in
{\leftskip .5in \rightskip .5in \noindent \ninerm \baselineskip=10pt
\bigskip\bigskip
Figure 1.
The contour $\CC^L_\vphi$.  The cut (wavy line) is oriented at an
angle $\vphi$ from the negative $y$-axis. The circle is at $|u| = 1$.
\smallskip}} \bigskip
\endinsert

The choice of the contours $\cL, \cR$ is dictated by the following
reasoning.  In planar quantization, the $x$-component of momentum
is $P_x = -\ma (u - 1/u )$.  Thus left-moving particles with $P_x <0$
have $u>1$ whereas right moving particles have $P_x > 0, ~ u<1$.
Thus the choice of $\cL$ versus $\cR$, i.e.
$|u| > 1$ versus $|u| < 1$, corresponds to an analytic continuation
of left versus right moving particles in planar quantization.
This choice can also be viewed as dictated by \eIIIxii.  The
precise orientation of the branch cut in figures 1,2 is chosen
for later convenience.

\midinsert
\epsfxsize = 2in
\bigskip
\vbox{\vskip -.1in\hbox{\centerline{\epsffile{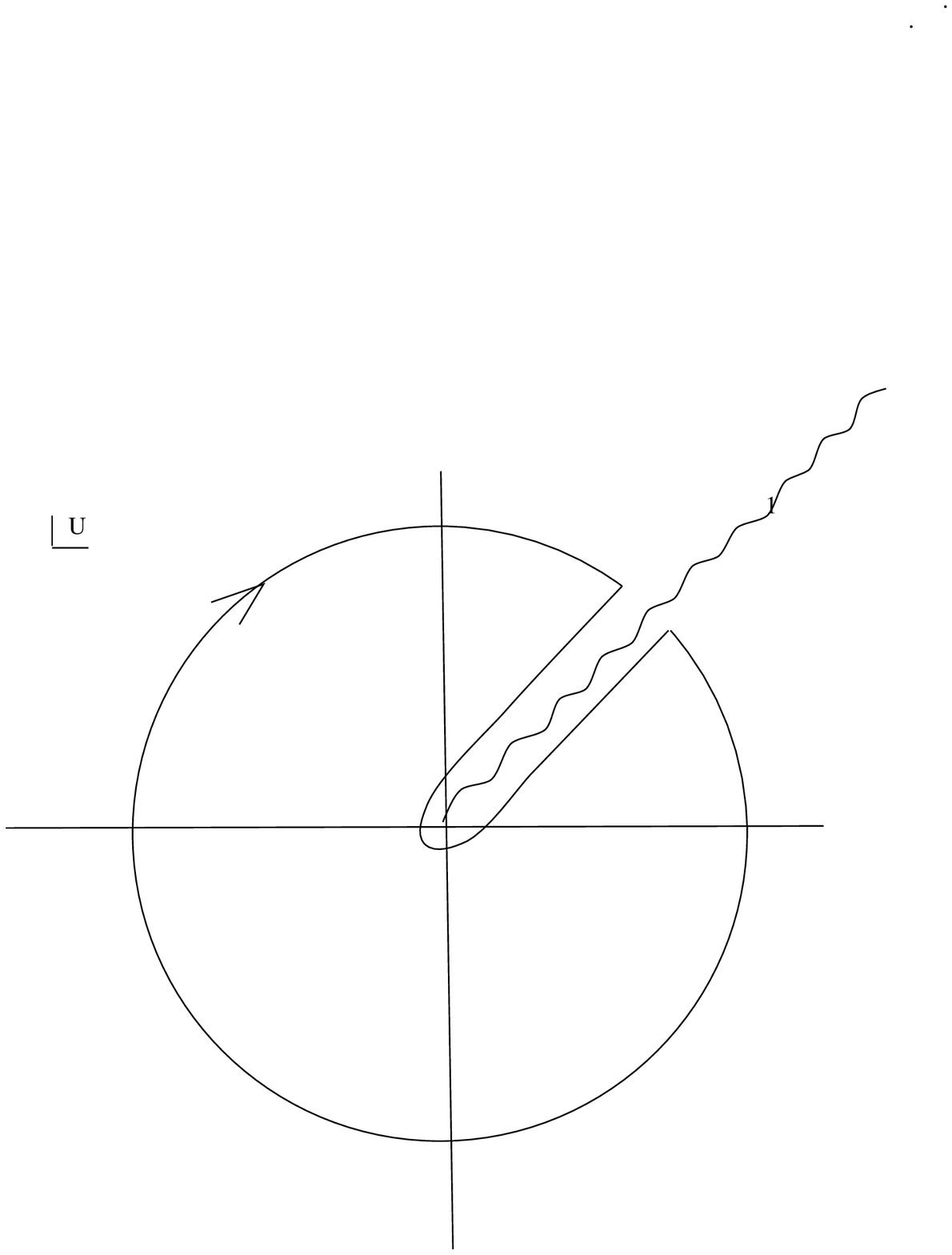}}}
\vskip .1in
{\leftskip .5in \rightskip .5in \noindent \ninerm \baselineskip=10pt
Figure 2.
The contour $\CC^R_\vphi$.  The cut  is oriented at an
angle $\vphi$ from the positive $y$-axis.  The circle is at
$|u| = 1$.
\smallskip}} \bigskip
\endinsert

The operators $\bpm (u) , ~ \bbpm (u)$ have the following analytic
expansions in $u$:
\eqn\eIIIxiv{\eqalign{
\bpm (u) &= \pm i \sum_{\om \in \Zmath } \Gamma (\ha - \om )
\> \ma^\om ~ b^\pm_\om ~ u^\om   \cr
\bbpm (u) &= \pm  \sum_{\om \in \Zmath } \Gamma (\ha - \om )
\> \ma^\om ~ \bb^\pm_\om ~ u^{-\om}  .    \cr }}
The Gamma functions in these expressions are included
to simplify subsequent expressions.  The contour integrals can
now be performed in \eIIxx\ with the replacement \eIIIxiii.  One
needs
\eqna\eIIIxv
$$\eqalignno{
e^{-i \om \vphi} ~ I_\om (\ma r) &= - \int_{\cR} \dua  ~ u^\om
{}~ \ez &\eIIIxv {a} \cr
e^{i \om \vphi} ~ I_\om (\ma r) &=  \int_{\cL} \dua  ~ u^{-\om}
{}~ \ez &\eIIIxv {b} , \cr}
$$
where $I_\om$ is the modified Bessel function\foot{The
formulas \eIIIxv\ are proven using
$$I_\om (r) = \oint_{\CC_0} \dua ~ u^\om ~ e^{ r(u+ 1/u )/2 } ,$$
where $\CC_0$ is the same as the contour $\cR$ rotated by an angle
$\vphi$ clockwise.  To prove \eIIIxv{b} one makes the change of
variable $u\to 1/u$ in \eIIIxv{a} and interchanges
$z \leftrightarrow \zb$.}.
The result is
\eqn\eIIIxvii{
\col{\psibpm}{\psipm} = \sum_{\om \in \Zmath}
\bpm_\om ~ \Psi^{(a)}_{-\om - 1/2} ~+~
\bbpm_\om ~ \bar{\Psi}^{(a)}_{-\om -1/2} , }
where
\eqn\eIIIxviii{\eqalign{
\Psi^{(a)}_{-\om -1/2} &=
\Gamma (\ha - \om ) ~ \ma^{\om + 1/2} ~
\col{i \> e^{i(\ha - \om )\vphi} ~ I_{\ha - \om} (\ma r) }
{e^{-i(\om + \ha )\vphi} ~ I_{-\om -\ha} (\ma r)}  \cr
\bar{\Psi}^{(a)}_{-\om - 1/2}
&=
\Gamma (\ha - \om ) ~ \ma^{\om + 1/2} ~
\col{ e^{i(\ha + \om )\vphi} ~ I_{-\ha - \om} (\ma r) }
{-i \> e^{-i(\ha - \om )\vphi} ~ I_{\ha - \om } (\ma r)}  .\cr}}
(Henceforth the labels $a,p$ represent anti-periodic versus periodic.)

One can verify explicitly that the expressions \eIIIxvii\
continue to satisfy the Dirac equations of motion \eIIxi\ using
the identities
\eqn\eIIIxix{
r\d_r \> I_\om (\ma r) \pm \om \> I_\om (\ma r)
= \ma r ~ I_{\om \mp 1} (\ma r) . }
In conventional approaches to radial quantization the expansions \eIIIxviii\
are found directly as solutions to the radial Dirac equation.

Similar but not as elegant results apply to the periodic sector.  Now
one takes the combination $\squ \> \bhpm (u)$ to be single valued, which
means that the expansions analagous to \eIIIxiv\ are in half-integral
powers of $u$.  It is possible to define a prescription for analytic
continuation as in \eIIIxiii, however the absence of a branch cut
in this situation does not lead to as clear a distinction between left
and right as for the anti-periodic sector.  The result is that
 now
\eqn\epercont{
\int_{-\infty}^\infty \du ~ \bhpm (u) \rightarrow
\( \oint_{\CC_<} \dua \( \bpm_< (u) + \bbpm_< (u) \)
+ \int_{\CC_> } \dua \( \bpm_> (u) + \bbpm_> (u) \) \) , }
where
\eqn\eperiodic{\eqalign{
\bpm (u) &= \pm i \sum_{\om \leq -1/2} \Ga (\ha - \om ) ~ \ma^\om ~
b^\pm_\om u^\om
{}~\pm~ \sum_{\om \geq 1/2} \frac{2\pi}{\Ga (\om + \ha ) }
(-1)^{\om + 1/2} ~ \ma^\om ~ b^\pm_\om u^\om  \cr
&\equiv \bpm_< (u) + \bpm_> (u) \cr
\bbpm (u) &= \pm  \sum_{\om \leq -1/2} \Ga (\ha - \om ) ~\ma^\om ~ \bbpm_\om
u^{-\om}
{}~\pm~ i \sum_{\om \geq 1/2} \frac{2\pi}{\Ga (\om + \ha ) }
(-1)^{1/2 -\om } ~ \ma^\om ~ \bbpm_\om u^{-\om}
\cr
&\equiv \bbpm_< (u) + \bbpm_> (u) . \cr }}
The contour $\CC_<$ is defined to be a closed contour on
the unit circle,
whereas $\CC_>$
 runs from $0$ to $\infty$ along a ray at an angle $\vphi$
above the negative $x$-axis in the $u$ plane.
Using the integrals
\eqn\eintper{\eqalign{
e^{-i n \vphi} ~ I_n (\ma r)  &= \oint_{\CC_<} \dua ~ u^n ~ \ez  , \cr
e^{-i n \vphi} ~ K_n (\ma r)  &= i \pi (-)^{n+1}
\int_{\CC_>} \dua ~ u^n ~ \ez  , \cr}}
one finds an expansion of the form \eIIIvii\ where now
the  sum runs over $\om \in \Zmath + 1/2$, and the basis spinors
differ from the expressions \eIIIviii\ only
for $\om\geq 1/2$:
\eqn\eIIIxix{\eqalign{
\Psi^{(p)}_{-\om -1/2} &= \frac{2 \ma^{\om + 1/2}}{\Ga (\ha + \om )}
{}~ \col{-i \> e^{i(\ha - \om )\vphi} ~ K_{ \om - \ha} (\ma r) }
{e^{-i(\om + \ha)\vphi} ~ K_{\om +\ha} (\ma r)} ~~~~\om \geq 1/2   \cr
\bar{\Psi}^{(p)}_{-\om -1/2} &= \frac{2 \ma^{\om + 1/2}}{\Ga (\ha + \om )}
{}~ \col{ e^{i(\ha + \om )\vphi} ~ K_{ \om + \ha} (\ma r) }
{i \> e^{i(\om - \ha)\vphi} ~ K_{\om -\ha} (\ma r)} ~~~~\om \geq 1/2
. \cr}}
In the above expansions the distinction between left and right is
as clear as in the anti-periodic sector.  Henceforth, unless
otherwise indicated the operators $\bpm (u) , ~ \bbpm (u)$ will be
in the anti-periodic sector.

The commutation relations of the operators $\bpm_\om, ~ \bbpm_\om$
can be computed from \eIIIv\ and the expansions \eIIIxvii\ in
either sector.  We now derive these same commutation relations in
a different way, one that emphasizes the momentum space operators.
Since the expansion of the fermion fields follows from an analytic
continuation of \eIIxx, the commutation relations of the $b$
operators are simply related to the commutation relations of the
$\bh$'s given in \eIIxix.  In the anti-periodic sector the relations
\eIIIv\ imply
\eqn\eIIIxx{\eqalign{
\{ \bp (u) , \bm (u') \} = 2 \pi^2 i  u' \, \de (u + u' ) , ~~&~~~
\{ \bbp (u) , \bbm (u') \} = - 2 \pi^2 i  u' \, \de (u + u' ) \cr
\{ b(u) , \bb (u') \} &= 0. \cr}}
One has the formal identity
\eqn\eIIIxxi{
2\pi i \> \de (u+u') = \inv{u'} ~ \sum_{n\in \Zmath} (-)^n \(
\frac{u}{u'} \)^n . }
This analytic $\de$-function is constructed to satisfy
\eqn\eIIIxxii{
\oint_0 du' ~ \de (u'-u) f(u' ) = f(u)  }
for any $f(u)$ with a Laurent series expansion about $u=0$.
Inserting the expansions \eIIIxiv\ into \eIIIxx, and using
the identity
\eqn\eIIIxxiii{
\Ga (\ha - \om ) \Ga (\ha + \om ) = \frac{\pi}{\cos (\pi \om )} , }
one finds
\eqn\eIIIxxiv{
\{ \bp_\om , \bm_{\om'} \} = \de_{\om , -\om'}
, ~~~~~\{ \bbp_\om , \bbm_{\om'} \} = \de_{\om , -\om'}
, ~~~~~\{ b_\om , \bb_{\om '} \} = 0. }
These commutation relations also hold in the periodic sector.
The operators $\bpm_\om , \bbpm_\om$ have scaling dimension $ D= -\om$,
and Lorentz spin $\pm \om$:
\eqn\eIIIxxiiib{
\[ L , \bpm_\om \] = -\om ~ \bpm_\om , ~~~~~
\[ L , \bbpm_\om \] = \om ~ \bbpm_\om , }
in accordance with \eIIIix.

As we now explain, the fock spaces built with the fermionic oscillators
correspond to the space of fields.  Consider first the periodic sector.
We define the physical vacuum as follows:
\eqn\eIIIxxv{\eqalign{
\bpm_\om \> \rvac &= \bbpm_\om \> \rvac = 0 , ~~~~~\om \geq 1/2  \cr
\lvac \> \bpm_\om &= \lvac \> \bbpm_\om = 0 , ~~~~~\om \leq -1/2 . \cr }}
Define the left and right periodic fock spaces:
\eqn\eIIIxxvi{\eqalign{
\CH^L_p &= \left\{ \bm_{-\om_1}
\bm_{-\om_2} \cdots \bp_{-\om_1 '} \bp_{-\om_2 '}
\cdots \rvac \right\} \cr
\CH^R_p &= \left\{ \bbm_{-\om_1}
\bbm_{-\om_2} \cdots \bbp_{-\om_1 '} \bbp_{-\om_2 '}
\cdots \rvac \right\} , \cr}}
where $\om , \om'  \geq 1/2$.
Let us first illustrate the explicit connection with the
space of fields by considering the fermion fields.  One needs the
asymptotic expansions of the Bessel functions
\eqn\eIIIxxvii{\eqalign{
I_\om (r) &= \sum_{k=0}^\infty \frac{1}{k! \> \Ga (\om + k +1 ) } ~
\( \frac{r}{2} \)^{\om + 2k}  \cr
K_n (r) &= \inv{2} \sum_{k=0}^{n-1} (-1)^k \frac{(n-k-1)!}{k!} \( \frac{r}{2}
\)^{2k-n}  \cr
&~~~~+ (-1)^{n+1} \sum_{k=0}^\infty \frac{(r/2)^{n+2k}}{k! (n+k)!}
\( \log (r/2) - \psi(k+1)/2 - \psi(n+k+1)/2 \) , \cr}}
where $\psi (x) = \d_x \log (\Ga (x) )$, and $n\geq 0$.
Using this, one finds
\eqn\eIIIxxviii{\eqalign{
\lim_{r\to 0} ~~ \psi_\pm (r, \vphi ) \rvac &=
\lim_{r\to 0} \sum_\om \( \bpm_\om ~ z^{-\om - 1/2} ~-~ \frac{i\ma}{(1/2-\om)}
{}~ \bbpm_\om ~ \zb^{1/2 - \om} \) \rvac \cr
&= \bpm_{-1/2} ~ \rvac . \cr }}
Similarly,
\eqn\eIIIxxix{
\lim_{r\to 0} ~~ \psibpm (r, \vphi ) ~ \rvac = \bbpm_{-\ha} ~ \rvac . }
Following the previously introduced terminology, the fields
$\psipm , ~ \psibpm $ are in $\CH^{(0)}_F$, and \eIIIx\ holds.
Indeed it is evident that the space $\CH^{(0)}_F$ is identical to
the space of fields in the massless limit from the
expressions:\foot{For
 a review of
the conformal field theory properties of free fermions and bosons
used in this
paper see \ref\rginsparg{\Ginsparg}.}
\eqn\eIIIxxx{
\col{\psibpm}{\psipm} ~ \mtoo ~ \sum_\om ~
\col{\bbpm_\om ~ \zb^{-\om-1/2} }{\bpm_\om ~ z^{-\om-1/2} } . }
The above equation is valid in either sector.
The higher modes correspond to derivatives of the fermions:
\eqn\eIIIxxxi{
\d_z^n \psipm (0) \rvac = n! ~ \bpm_{-n-\ha} \rvac , ~~~~~
\d_\zb^n \psibpm (0) \rvac = n! ~  \bbpm_{-n-\ha} \rvac . }
Mixed derivative fields can always be simplified using
the equations of motion to relate them linearly to the above
states and are thus not in $\CH^{(0)}_F$.  For example
\eqn\eIIIxxxii{
\d_\zb^m \d_z^{n+m} \psipm (0) \rvac = \ma^{2m} n! ~ \bpm_{-n-\ha} \rvac . }
Other states in $\CH^{L,R}_p$ correspond to composite operators.
For example, consider the $\uone$ current $J_z = \psip \psim $,
$J_\zb = \psibp \psibm $.  One has
\eqn\eIIIxxxiii{
J_z (0) \rvac = \bp_{-\ha} \> \bm_{-\ha} \> \rvac ,
{}~~~~~J_\zb (0) \rvac = \bbp_{-\ha} \> \bbm_{-\ha} \> \rvac . }

We now turn to the anti-periodic sector.  Due to the existence of the
zero modes $\bpm_0 , \bbpm_0$, the `vacuum'
in this sector is doubly
degenerate for both left and right.  Define these vacua as
$\rvacpm_L$ and $\rvacpm_R$, characterized by
\eqn\eIIIxxxiv{\eqalign{
\bpm_0 \> \va{\mp \ha}_L = \rvacpm_L , ~~~~ \bpm_0 \> \rvacpm_L = 0 ,
{}~~~~ \bpm_n \> \rvacpm_L = 0,~~~~&n\geq 1 \cr
\bbpm_0 \> \va{\mp \ha}_R = \rvacpm_R , ~~~~ \bbpm_0 \> \rvacpm_R = 0 ,
{}~~~~ \bbpm_n \> \rvacpm_R = 0,~~~~&n\geq 1 . \cr}}
The vacua $\lvacpm$ are defined by the inner products
\eqn\eIIIxxxv{
{}_L \lva{\mp \ha} \pm \ha \rangle_L = {}_R \langle \mp \ha \rvacpm_R = 1.}
These vacuum states have $U(1)$ charge $\pm 1/2$.  The anti-periodic
fock spaces are defined as
\eqn\eIIIxxxvb{\eqalign{
\CH^L_{a_{\pm}}
&= \left\{ \bm_{-n_1} \bm_{-n_2} \cdots \bp_{-n_1 '} \bp_{-n_2 '}
\cdots \rvacpm_L \right\} \cr
\CH^R_{a_{\pm}}
&= \left\{ \bbm_{-n_1} \bbm_{-n_2} \cdots \bbp_{-n_1 '} \bbp_{-n_2 '}
\cdots \rvacpm_R \right\} , \cr}}
for $n , n' \geq 1$.

The fock states in $\CH^{L,R}_{a_{\pm}}$, including the vacuum
states, should also be identified with fields.
In order to identify fields corresponding to $\rvacpm_{L,R}$, we
first study the conformal limit.  In this limit the Dirac
theory is exactly equivalent to a massless scalar field $\phi$,which
splits into chiral pieces $\phi = \vphi^L (z) + \vphi^R (\zb )$.
 From the bosonized expressions for the fermion fields \eIIix,
one finds the following operator product expansions:
\eqn\eIIIxxxvi{\eqalign{
\psipm (z) ~ e^{\mp i \vphi^L (0)/2 } ~ &\sim ~ \inv{\sqrt{z}} ~
e^{\pm i \vphi^L (0)/2 }  + \ldots \cr
\psibpm (\zb) ~ e^{\pm i \vphi^R (0)/2 } ~ &\sim ~ \inv{\sqrt{\zb}} ~
e^{\mp i \vphi^R (0)/2 } + \ldots  \cr}}
Defining
\eqn\eIIIxxxvii{
e^{\pm i \vphi^L (0) /2 } \> \rvac = \rvacpm^{\ma = 0}_{L} ,
{}~~~~~e^{\mp i \vphi^R (0)/2 } \> \rvac = \rvacpm^{\ma =0}_R , }
one has
\eqn\eIIIxxxviii{
\bpm_0 \> \va{\mp \ha}^{\ma = 0}_{L} = \oint_0 \dz ~  \inv{\sqrt{z}} ~
\psipm (z) ~ e^{\mp i \vphi^L (0)/2} \> \rvac = \rvacpm^{\ma = 0}_L , }
and similarly for $\bbpm_0 \va{\mp \ha}_R$.
 From \eIIIxxxvi\ one has the well-known result
that the fields corresponding to the
states $\rvacpm^{\ma = 0}_{L,R}$ in the conformal field theory
are the analog of the Ising
disorder and spin fields for this $U(1)$ theory.

It is essential in what we are doing that the states $\rvacpm_{L,R}$ were
precisely defined in the {\it massive} theory.  Based on the above massless
reasoning, we propose the following identification:
\eqn\eIIIxxxix{
e^{\pm i \phi (0)/2 } \> \rvac ~\propto~ \( \rvacpm_L \ot \rvacmp_R \)
\equiv \rvacpm , }
where $\phi (z, \zb )$ is the local SG field.
The correct interpretation of \eIIIxxxix\
is that it provides a non-perturbative {\it definition} of the LHS, and
furthermore, a quasi-chiral factorization of this field.
Note that with this identification, $e^{\pm i \phi (0)/2 }$ is
$\uone$ neutral.  As in the conformal field theory, the states
$\rvacpm_{L,R}$ have the following Lorentz spin and dimension:
\eqn\eIIIxxxx{\eqalign{
L\>  \rvacpm_L = 1/8  ~ \rvacpm_L , ~~~~~&~~~~~
L \> \rvacpm_R = - 1/8 ~ \rvacpm_R  \cr
D \> \rvacpm_{L,R}  &= 1/8 ~ \rvacpm_{L,R} . \cr}}
These formulas will be justified in the massive theory in section 6.
The identification \eIIIxxxix\ implies that $e^{\pm i \phi (0)/2}$ has
scaling dimension $D=1/4$ and is Lorentz spinless.

Other examples of field identifications are
\eqn\eIIIxxxxi{\eqalign{
b^+_{-1} \va{- \ha }_L &= \d_z \( e^{i \vphi^L (0) /2} \) \rvac \cr
b^+_{-1} \va{+ \ha }_L & =  : \psi_+ (0) \> e^{i\vphi^L (0)/2} :
\rvac = e^{3i\vphi^L (0)/2 } \> \rvac . \cr}}
The normal ordering in the above equation means a regularized product
of the two fields.

The spaces $\CH^{L,R}_F$ can also be obtained directly from
the $\bpm (u) , \bbpm (u)$ operators.  For example, in either sector
one has
\eqn\eIIIxxxxib{\eqalign{
\lim_{u\to \infty} ~ (\ma u )^\om ~ \bpm (u) ~ \va{\ep}
&= \pm i ~ \Ga (\om + \ha ) ~ \bpm_{-\om} ~ \va{\ep} \cr
\lim_{u\to 0} ~ \( \frac{\ma}{u} \)^\om ~ \bbpm (u) ~ \va{\ep}
&= \pm  ~ \Ga (\om + \ha ) ~ \bbpm_{-\om} ~ \va{\ep} , \cr}}
where $\ep = 0, \pm 1/2$, depending on the sector.  This illustrates
eq. \eIIIxii.

Correlation functions in the periodic sector are easily computed
in this framework.  For example,
\eqn\eIIIxxxxii{
\lvac \psi_- (z , \zb ) ~ \psi_+ (0) \rvac ~=~
\lvac \> \psi_- (z , \zb ) ~ b^+_{-\ha} \> \rvac
{}~=~ \( \Psi^{(p)}_{-1} \)_2 , }
where
$ \( \Psi^{(p)}_{-1} \)_2$ is the second component of the spinor
$\Psi^{(p)}_{-1}$.  In this way one reproduces the results
\eIIxxxx.  Similarly one obtains correlation functions
of the $\uone$ current:
\eqn\eIIIxxxxiii{\eqalign{
\lvac \> J_z (z , \zb ) \> J_z (0) \> \rvac &= 4 \ma^2 \> \frac{\zb}{z} \>
\( K_1 (\ma r)\)^2  \cr
\lvac \> J_\zb (z , \zb ) \> J_\zb (0) \> \rvac &= 4 \ma^2 \> \frac{z}{\zb} \>
\( K_1 (\ma r)\)^2  \cr
\lvac \> J_\zb (z , \zb ) \> J_z (0) \> \rvac &= - 4 \ma^2 ~
\( K_0 (\ma r)\)^2  . \cr }}

Somewhat less trivial correlation functions are those of the
fermions in the anti-periodic sector:
\eqn\eIIIxxxxiv{\eqalign{
\lva{\ha} ~ \psi_+ & (r, \vphi ) ~ \psi_- (r' , \vphi ' ) ~
\va{-\ha}_{\vphi = \vphi' = 0}   \cr
& =~ \ma \pi \[ \( \sum_{n=1}^\infty (-)^n ( I_{-n-\ha} (\ma r ) I_{n-\ha}
(\ma r' ) - I_{\ha-n} (\ma r) I_{\ha + n} ( \ma r' )) \)
- I_\ha (\ma r) I_\ha (\ma r' ) \] \cr
& =~ \mtoo ~ \sqrt{\frac{z}{z'}} \> \inv{z-z'} . \cr }}

\bigskip
\newsec{Momentum Space Correlation Functions and Form Factors}
\bigskip

The analytic properties of the operators $\bpm (u) , ~ \bbpm (u)$
acquired in radial quantization lead to well defined vacuum
expectation values of products of such operators. As we now show,
these vacuum expectation values have a physical meaning as form-factors.
The properly normalized one particle states with $\uone$ charge $\pm 1$
are
\eqn\eIVi{
\eqalign{
\va{\uparrow, \th } = \inv{2\pi} ~ d^\dag (\th ) \rvac , ~~~&~~~
\va{\downarrow , \th } = \inv{2\pi} ~ c^\dag (\th ) \rvac \cr
\lva{\downarrow , \th '} \uparrow , \th \rangle &= \de (\th - \th' ) .\cr}}
Consider the following 2-particle form-factor in the usual planar
quantization
\eqn\eIVii{\eqalign{
f_{\pm \ha} (\th , \th ' ) &=  \lvac ~e^{ \pm i \phi (0)/2 } ~
\va{\uparrow, \th ; \downarrow , \th' } \cr
&= \inv{ 4 \pi^2 i } ~ \lvac e^{\pm i \phi (0) /2} ~ \bhp (u)
\bhm (u' ) \rvac , \cr }}
where as usual $u = e^\th$.  This form-factor is well-defined in the
SG theory, where $\phi$ is the SG field and $\va{\uparrow , \th ;
\downarrow , \th' } $ is a 2-soliton state.  Due to the expressions
\eIIxx, this form factor enters into the  computation of the
the correlation function
$$ \lvac e^{ \pm i \phi (0)/2 } ~ \psip (z , \zb ) ~ \psim (w , \bar{w} )
{}~\rvac .$$
In radial quantization this correlation function
is computed by analytically continuing the $u$-integrals in \eIIxx
{}~ to obtain an expression of the form \eIIIxxxxiv.
 This implies that the form-factors are computable in radial
quantization as follows:
\eqn\eIViii{
\eqalign{
\lvac ~ e^{\pm i \phi (0) / 2} ~ &\bhp (u) \bhm (u' ) ~ \rvac \cr
& = {}_L \lvacmp b^+ (u) b^- (u' ) ~ \rvacpm_L = {}_R \lvacpm
{}~ \bb^+ (u) \bb^- (u' ) ~\rvacmp_R . \cr }}

We now describe some techniques for computing vacuum expectation values
such as \eIViii. Based on the vacuum properties \eIIIxxxiv, we define
the following normal ordering prescription:
\eqn\eIViv{
: \bpm_n \> \bmp_{-n} : = - \bmp_{-n} \> \bpm_n    ,
{}~~~~~: \bbpm_n \> \bbmp_{-n} : = - \bbmp_{-n} \> \bbpm_n    ,
{}~~~~~n > 0. }
To simplify the computation of vacuum expectation values, it proves
useful to separate out the zero modes:
\eqn\eIVv{\eqalign{
\bpm (u) &=  \pm i \sqrt{\pi} ~ \bpm_0 ~+~ \bpm_\bullet (u) \cr
\bbpm (u) &=  \pm  \sqrt{\pi} ~ \bbpm_0 ~+~ \bbpm_\bullet (u) .\cr}}
One has
\eqn\eIVvi{\eqalign{
\bbul^\pm (u) \> \bbul^\mp (u' ) &= : \bbul^\pm (u) \> \bbul^\mp (u' ):
 ~ - ~\pi  ~ \frac{u}{u+u' }  \cr
\bbbul^\pm (u) \> \bbbul^\mp (u' ) &= : \bbbul^\pm (u) \> \bbbul^\mp (u' ):
 ~ + ~\pi  ~ \frac{u'}{u+u' }  . \cr}}
 The Wick theorem may now be used with the pieces $\bbul^\pm (u) , ~
 \bbbul^\pm (u) $.  The two point functions are easily computed to be
 \eqn\eIVvii{\eqalign{
 {}_L \lvacm ~ b^+ (u) \> b^- (u' ) ~ \rvacp_L
 = {}_R \lvacp ~ \bb^+ (u) \> \bb^- (u' ) ~ \rvacm_R  &= \pi
 \frac{u'}{u+u'} \cr
 {}_L \lvacp ~ b^+ (u) \> b^- (u' ) ~ \rvacm_L
 = {}_R \lvacm ~ \bb^+ (u) \> \bb^- (u' ) ~ \rvacp_R  &= - \pi
 \frac{u}{u+u'} . \cr}}

The form-factors $f_{\pm \ha} (\th , \th' )$ as computed from
\eIViii\ and \eIVvii\ agree exactly with the known result\ref\rmss{\MSS}\rform.
As we will show more easily in section 6, this result extends to the
multiparticle form factors.

\bigskip
\newsec{The Spectrum Generating Affine Lie Algebra}
\bigskip
\noindent
5.1 {\it ~~ Quasi-Chiral Splitting of the Conserved Charges}
\medskip

\def\Psib{\bar{\Psi}}

The quasi-chiral factorization described above leads to some additional
structures for the conserved charges.  In radial quantization, given
a conserved current $J_z , ~ J_\zb$, the conserved charge is
\eqn\eVi{
Q = \inv{4\pi} \int_{-\pi}^\pi ~ r \> d\vphi
\( e^{i\vphi} \>  J_z  ~+~ e^{-i\vphi} \>  J_\zb \) . }
All of the conserved charges constructed in section 2 can thereby be
expressed in terms of the radial modes $b^\pm_\om , ~ \bb^\pm_\om$.
More specifically, define the inner product of two spinors
$A = \col{\bar{a}}{a}$, $B = \col{\bar{b}}{b}$ as
\eqn\eVii{
(A,B) = \inv{4\pi} \int_{-\pi}^\pi ~ r d\vphi
\( e^{ i\vphi} ~ a \>  b ~+~ e^{-i\vphi} ~ \bar{a} \> \bar{b} \) . }
Using the Wronskian identities
\eqn\eViii{\eqalign{
I_\om ( r ) ~ K_{\om + 1 } ( r ) ~+~ I_{\om + 1} ( r ) ~ K_{\om} (r)
&= \inv{r} \cr
I_\om ( r ) ~ I_{-\om - 1 } ( r ) ~-~ I_{\om + 1} ( r ) ~ I_{-\om} (r)
& = -2 \> \frac{\sin (\pi \om )}{\pi r } , \cr }}
one finds simple inner products among the spinors in the radial
basis.  In either sector:
\eqn\eViv{
\( \Psi_\om , \Psi_{-\om ' -1} \)
= \( \Psib_\om , \Psib_{-\om ' -1} \)  = \de_{\om , \om ' } ,
{}~~~~~\( \Psi_\om , \Psib_{\om ' } \) = 0 . }
The conserved charges can all be expressed using the above inner
product, e.g.
\eqn\eVv{\eqalign{
Q^\pm_{-1} = \inv{2}  \( \Psi_\pm , \d_z \Psi_\pm \)& , ~~~~~
Q^\pm_{1} = \inv{2} \( \Psi_\pm , \d_\zb \Psi_\pm \) \cr
& T_0 = ~: \( \Psi_+ , \Psi_- \) :  , \cr }}
and are thus easily expressed in terms of the $b, \bb$-modes.

It is more useful for us to work in momentum space, i.e. with the
analogs of \eIIxxi\ in radial quantization.  The proper analytic
continuation of \eIIxxi\ is the following:
\eqn\eVvi{
\int_{-\infty}^\infty \du ~ \bh (u) ~ \bh (-u ) \to
2 \oint \dua  ~ b (u) ~ b (-u )
-2 \oint \dua ~ \bb (u) \bb (-u ) . }
One can check explicitly that the prescription \eVvi\ is equivalent
to the position space expressions \eVv.  Thus, the conserved charges
split into left and right pieces:
\eqn\eVvii{\eqalign{
Q^\pm_n ~&=~ Q^{\pm , L}_n ~+~ Q^{\pm ,R}_{-n}  \cr
\al_n ~&=~ \al^L_n ~+~ \al^R_{-n} \cr
L ~&=~ L^L ~+~ L^R . \cr }}
The additional minus sign in the subscript $-n$ of the right piece
of the charges in comparison to the left piece is dictated by Lorentz
covariance.
The explicit expressions are as follows:
\eqn\eVviii{\eqalign{
\al^L_n &= \frac{\ma^{|n|}}\pi ~ \oint \dua ~ u^{-n} ~ : b^+ (u) ~b^- (-u): \cr
&= \frac{\ma^{|n| + n}}{\pi} ~ \sum_\om (-)^{n-\om} \Ga (\ha - \om )
\Ga (\ha + \om -n ) ~ : b^+_\om ~ b^-_{n-\om } : \cr
\al^R_n &= - \frac{\ma^{|n|}}\pi ~ \oint \dua ~ u^{n} ~ : \bb^+ (u)
{}~\bb^- (-u): \cr
&= \frac{\ma^{|n| + n}}{\pi} ~ \sum_\om (-)^{n+\om} \Ga (\ha - \om )
\Ga (\ha + \om -n ) ~ : \bb^+_\om ~ \bb^-_{n-\om } : \cr
Q^{\pm ,  L}_n &= \frac{\ma^{|n|}}{2 \pi} ~ \oint \dua ~ u^{-n} ~  b^\pm (u)
{}~b^\pm  (-u) \cr
&= \frac{\ma^{|n| + n}}{2\pi} ~ \sum_\om (-)^{n-\om + 1} \Ga (\ha - \om )
\Ga (\ha + \om -n ) ~  b^\pm_\om ~ b^\pm_{n-\om }  \cr
Q^{\pm ,  R}_n &= - \frac{\ma^{|n|}}{2\pi} ~ \oint \dua ~ u^{n} ~  \bb^\pm (u)
{}~\bb^\pm (-u) \cr
&= \frac{\ma^{|n| + n}}{2\pi} ~ \sum_\om (-)^{n-\om + 1} \Ga (\ha - \om )
\Ga (\ha + \om -n ) ~  \bb^\pm_\om ~ \bb^\pm_{n-\om }  \cr
L^L &= \inv{\pi} ~ \oint \dua ~ : b^+ (u) u \d_u b^- (-u ):
{}~ = -  \sum_\om ~ \om ~ : b^+_\om \> b^-_{-\om} :  \cr
L^R &= - \inv{\pi} ~ \oint \dua ~ : \bb^+ (u) u \d_u \bb^- (-u ):
{}~ =   \sum_\om ~ \om ~ : \bb^+_\om \> \bb^-_{-\om} :  ~ . \cr}}
The above expressions were derived in the anti-periodic sector, but
similar (but not identical) expressions apply to the periodic sector.
Namely,
if one first expresses the product of
Gamma-functions as a rational function of $\om$ using \eIIIxxiii, and then
lets the sum run over $\om \in \Zmath + 1/2$ while omitting terms that
are singular one obtains the correct periodic sector expressions.

The operators we have introduced have the following Lorentz
spin properties
\eqn\eVix{
\eqalign{
\[ L^L , \bpm_\om \] &= -\om ~ \bpm_\om , ~~~~~
\[ L^R , \bbpm_\om \] = \om ~ \bbpm_\om  \cr
\[ L^L , \CO_n^L \] &= -n  ~  \CO^L_n , ~~~~~
\[ L^R , \CO^R_n \] = n ~ \CO^R_n  ,\cr}}
where $\CO_n^L = Q^{\pm ,L}_n , {\rm or} ~ \al^L_n$, and similarly
for $\CO^R_n$.  The scaling operator is realized as
\eqn\eVx{
D = L^L - L^R + \ma \d_\ma , }
so that
\eqn\eVxi{
\eqalign{
\[ D , \bpm_\om \] = -\om ~ \bpm_\om , ~~~&~~~
\[ D , \bbpm_\om \] = - \om ~ \bbpm_\om  \cr
\[ D , \CO^{L,R}_n \] &= |n| ~ \CO^{L,R}_n . \cr }}
Here we see an important distinction with the situation encountered
in conformal field theory, in that all the operators $\CO^{L,R}_n$
have positive scaling dimension.  Note that the relations
\eVvii,\eVix\ reproduce the unsplit relations \eIIxxv.

\def\ol#1{{\CO^L_{#1}}}
\def\or#1{{\CO^R_{#1}}}

We now compute the commutation relations of the left and right
components among themselves in the anti-periodic sector.
Obviously, any left operator commutes with a right operator.
This computation is difficult if one tries to use the
integrated expressions in \eVviii\ due to the infinite sums, so
we develop a different technique.  More generally, let
$\CO^{L,R}_n$ have contour integral expressions in momentum
space of the form
\eqn\eVxii{
\ol n = \oint \dua ~ u^{-n} ~ \CO^L (u) , ~~~~~~
\or n = \oint \dua ~ u^n ~ \CO^R (u) . }
The non-trivial contributions to $[\ol n , \ol m ]$ arise from
pole singularities in the product $\CO^L (u) \CO^L (u')$.
There is no translation invariance in $u$-space, so these poles can
be at either $u = \pm u'$, the poles being related to each other
by `crossing symmetry' $\th \to i\pi + \th$.  The operators
$\CO^L (u)$ and $\CO^R (u)$ have a development in opposite powers
of $u$, which can be seen from \eIIIxii, or  for example in \eIIIxiv.
Thus in the vacuum expectation value, $\CO^L (u) \CO^L (u' )$ is
properly defined as an expansion in powers of $u/u'$  for $u<u'$,
whereas $\CO^R (u) \CO^R (u' )$ is a well-defined expansion in powers
of $u'/u$ for $u>u'$.  Taking all of this into account, one has
\eqn\eVxiii{\eqalign{
\[ \ol n , \ol m \] &= \oint_{\pm u}
\frac{du'}{2\pi i u'} ~
\oint_0 \dua ~ (u')^{-m} u^{-n} ~ \CO^L (u) \CO^L (u') \cr
\[ \or n , \or m \] &= \oint_{\pm u'} \dua ~ \oint_0
\frac{du'}{2\pi i u'} ~
(u')^{m} u^{n} ~ \CO^R (u) \CO^R (u') , \cr}}
where in the first equation the $u'$ contour is larger than the $u$-contour,
and visa-versa for the second equation.

It is useful to introduce the concept of an operator product
expansion in momentum space simply defined as a way of displaying the
pole singularities in the product of the momentum space operators.
For the actual computation we will perform, we need the vacuum expectation
values
\eqn\eVxiv{\eqalign{
{}_L \lvacmp ~ :b^+ (u) b^- (-u): : b^+ (u') b^- (-u'): \rvacpm_L
&= \pi^2 \frac{u u'}{(u-u')^2} \cr
{}_L \lvacmp ~ b^+ (u) b^+ (-u)~   b^- (u') b^- (-u'): \rvacpm_L
&= \pi^2  u^2 \( \inv{(u-u')^2} ~-~ \inv{(u+u')^2} \), \cr }}
which are easily derived using the techniques of the last section.
Using this, one finds the following operator product expansions:
\eqn\eVxv{\eqalign{
:b^+ (u) b^- (-u) :~ : b^+ (u') b^- (-u'): ~&\sim~ \pi^2 \frac{u\> u'}{(u
-u')^2} \cr
:b^+ (u) b^- (-u): (b^\pm (u') b^\pm (-u') ) ~ &\sim ~
\pm \( \frac{\pi u}{u'-u} ~ b^\pm (u) b^\pm (-u') ~+~
\frac{\pi u}{u+u'} ~ b^\pm (u) b^\pm (u') \) \cr
(b^+ (u) b^+ (-u))~ (b^- (u') b^- (-u') ) ~&\sim~
\pi^2  u^2 \> \( \inv{(u-u')^2} - \inv{(u+u')^2} \) \cr
&~~~~+ \frac{\pi u}{u' -u} \( b^+ (u) b^- (-u') + b^+ (-u) b^-(u') \) \cr
&~~~~+ \frac{\pi u}{u' +u} \( b^+ (u) b^- (-u') + b^+ (-u) b^-(-u') \)
. \cr}}

Define as before
\eqn\eVxvi{
P_n^{L,R} = \al^{L,R}_n , ~~~~n ~~{\rm odd}; ~~~~~~~~~~~~
T_n^{L,R} = \al^{L,R}_n , ~~~~n ~~{\rm even}. }
Then using the above formulas one finds the following result:
\eqn\eVxvii{\eqalign{
\[ P^L_n , P^L_m \] &=  n ~ \ma^{2|n|} ~ \de_{n, -m}  \cr
\[ P^L_n , T^L_m \] &= \[ P^L_n , Q^{\pm,L}_m \] = 0  \cr
\[ T^L_n , T^L_m \] &= n~ \ma^{2|n|} ~ \de_{n, -m}  \cr
\[ T^L_n , Q^{\pm ,L}_m \] &= \pm 2 ~ \mn ~ Q^{\pm, L}_{n+m}   \cr
\[ Q^{+,L}_n , Q^{-,L}_m \] &= \mn ~ T^L_{n+m} ~+~
\frac{n}{2} ~ \ma^{2|n|} ~ \de_{n,-m}  . \cr}}
In deriving the last equation above, we have defined
\eqn\eVxviii{
:b^+_0 b^-_0 : \equiv \frac{1}{2} (b^+_0 b^-_0 - b^-_0 b^+_0 ) = b^+_0 b^-_0
-1/2 . }
This shifts $T^L_0 \to T^L_0 + 1/2$, so that
\eqn\eVxix{
T^L_0 ~ \rvacpm_{L} = \pm \ha ~ \rvacpm_L . }
This shifts the form of the central term from being proportional
to the value $(n-1)$ obtained from the contour integration to $n$.

The operators $T^L_n , ~ Q^{\pm , L}_n$ thus satisfy a level $1$
$\slh$ algebra, and the $P^L_n$ satisfy an infinite Heisenberg algebra.
A similar computation yields precisely the same algebra for
$P^R_n , T^R_n , ~ Q^{\pm , R}_n $, with the same level $1$.
The appearance of central terms is entirely consistent with
the results of section 2.  Namely, the unsplit charges in
\eVvii\ continue to satisfy the level $0$ algebra \eIIxxiii,
since the central terms cancel between left and right in
the computation $[\CO_n , \CO_m ] = [\ol n , \ol m ] +
[ \CO^R_{-n} , \CO^R_{-m} ] $.  The above results are for the
anti-periodic sector; the periodic sector will be considered
at the end of this section.

\bigskip
\noindent 5.2 {\it ~~Highest Weight Representations and the Space of Fields}
\medskip

Let us denote the $P_n$ extension of the algebra $\slh$ defined in
\eVxvii\ as $\slhh$.  As we have seen, the symmetry algebra factorizes
into $\slhh_L \ot \slhh_R$.  We now show that the algebra $\slhh$ is
a complete spectrum generating algebra for the anti-periodic sector, namely,
that the complete spectrum of quasi-chirally factorized fields can be
obtained from infinite highest weight representations of $\slhh$.

We first review the known structure of highest weight representations of
the algebra $\slh$\rkac.  At level $1$, there are only 2 highest weight
states $\va{\La_j}$, $j = 0,1/2$, satisfying\foot{We are
using the twisted basis defined in section 2.  The highest weight conditions
below were obtained from the more familiar relations
$j^a_{n>0} \va{\La_j} = 0$, $j^+_0 \va{\La_j} = 0$,
$j^0_0 \va{\La_j} = 2j \va{\La_j}$.}
\eqn\eVxx{\eqalign{
t^\pm_n \> \va{\La_j} &= t^0_n \> \va{\La_j} =0
{}~~~~~n\geq 1  \cr
t^0_0 \> \va{\La_0} = -\ha ~ \va{\La_0} ,~~~~&~~~~
t^0_0 \> \va{\La_\ha} = \ha ~ \va{\La_\ha} . \cr}}

Consider first the anti-periodic sector.
Using the expressions in \eVviii, one finds
\eqn\eVxxb{\eqalign{
Q^{\pm , L}_n ~ \rvacpm_L &= T^L_n ~ \rvacpm_L = P^L_n \rvacpm_L = 0,
{}~~~~~n\geq 1 \cr
T^L_0 ~ \rvacpm_L &= \pm \ha ~ \rvacpm_L . \cr}}
Comparing with \eVxx, one sees that the states $\rvacpm_L$ are
highest weight for the $\slh$ subalgebra of $\slhh$, with the
identification\foot{We remark that this is already rather
different than the situation in conformal current algebra, where
$L^L \va{\La_0} = 0$, $L^L \va{\La_\ha} = {\sc {\inv{4}}}  \va{\La_\ha}$.
Here the $L^L$ eigenvalues are both $1/8$.  It is the twist that
makes this possible.}
\eqn\eVxxi{
\rvacp_L = \va{\La_\ha} , ~~~~~\rvacm_L = \va{\La_0} . }

We only need to proceed to a few more levels to get a clear
picture.  One finds
\eqn\eVxxii{
Q^{\pm , L}_{-1} \rvacpm_L = 0, ~~~~~~~Q^{\pm , L}_{-1} ~
\rvacmp_L = -\ha ~ b^\pm_{-1} ~ \rvacpm_L . }
The first equation is a `null-state' condition.  The second equation
indicates that the spaces $\CH^L_{a_+}$ and $\CH^L_{a_-}$ mix
under the action of $\slh$.  Another important point is that one
cannot obtain all of the states in $\CH^L_a \equiv \CH^L_{a_+} \oplus
\CH^L_{a_-} $ from the action of $\slh$ alone.  One must also descend
with the $P_n$'s.  For example
\eqn\eVxxiii{
P^L_{-1} ~ \rvacpm_L = \inv{2} ~ b^\pm_{-1} \rvacmp_L , }
and this is the only way to obtain the state on the RHS.

With the above qualitative picture, one can now proceed to precisely
identify the spaces $\CH^{L,R}_a$ with $\slhh$ modules.  We do this
by introducing generating functions, or characters, for these spaces.
Let $q$ denote a formal parameter (not to be confused with the
$q$-deformation parameter). Using results familiar from the known
partition functions of free fermions\rginsparg, one has
\eqn\eVxxiv{\eqalign{
\tr{\CH^L_{a_\pm}}  \( q^L \) &= q^{1/8} ~ \prod_{n=1}^\infty
\( 1+q^n \)^2  \cr
\tr{\CH^R_{a_\pm}}  \( q^L \) &= q^{-1/8} ~ \prod_{n=1}^\infty
\( 1+q^{-n} \)^2  . \cr}}
(The power of $2$ comes from the fact that we have a complex fermion.)
In computing the above trace over $\CH^L$ ($\CH^R$) only
$L^L$ ($L^R$) matters in the sum $L=L^L + L^R$.
Also,
\eqn\eVxxv{
\tr{\hal} \( q^L \) = \tr{\halp} \( q^L \) + \tr{\halm} \( q^L \) , }
and similarly for $\har$.
We present explicit formulas for the left sector; identical
results apply in the right sector with everywhere $L\to R, ~q\to q^{-1}$.

Define the $\slhh_L$ modules $\vhh^{{}_L}_{\La_j}$ as follows,
\eqn\eVxxvi{\eqalign{
\vhh^{{}_L}_{\La_0} &\equiv \left\{ Q^{\pm,L}_{-n_1} ~Q^{\pm , L}_{-n_2} \cdots
T^L_{-n_1 '} T^L_{-n_2 '} \cdots P^L_{-n_1 ''} P^L_{-n_2 ''} \cdots
\rvacm_L \right\} \cr
\vhh^{{}_L}_{\La_\ha}
&\equiv \left\{ Q^{\pm,L}_{-n_1} ~Q^{\pm , L}_{-n_2} \cdots
T^L_{-n_1 '} T^L_{-n_2 '} \cdots P^L_{-n_1 ''} P^L_{-n_2 ''} \cdots
\rvacp_L \right\} , \cr}}
for $n,n' , n'' \geq 1$.  Define also
\eqn\eVxxvii{
\chihh^{{}_L}_j \equiv \tr{\vhh^{{}_L}_{\La_j}} ~ \( q^L \) . }
We will prove that
\eqn\eVxxviii{
\tr{\hal} \( q^L \) = \chihh^{{}_L}_0 ~+~ \chihh^{{}_L}_\ha . }
The above result implies that we have the identification
\eqn\eident{
\hal = \vhh^{{}_L}_{\Lambda_0} \oplus \vhh^{{}_L}_{\Lambda_\ha} . }

To compute $\chihh^{{}_L}_j$, note first that since the $P^L_n$'s commute
with the $\slh_L$ subalgebra of $\slhh_L$, one has the factorization
\eqn\eVxxix{
\chihh^{{}_L}_j = q^{1/8} ~ \chi^L_P ~ \chih^{tw,L}_j , }
where $\chi^L_P$ is the contribution from the Heisenberg algebra of
the $P^L_n$.
The latter contribution is easily seen to be
\eqn\eVxxx{
\chi^L_P = \prod_{n=1}^\infty ~ \inv{ (1-q^{2n-1} )} . }
The functions $\chih^{tw, L}_j$ are the twisted characters for
the $\slh$ algebra, which are defined and computed in the appendix.
The result is
\eqn\eVxxxi{
\chih^{tw,L}_0 = \chih^{tw,L}_\ha = \prod_{n=1}^\infty ~
\inv{(1-q^{2n-1} )} . }
Putting these pieces together, to establish the main result \eVxxviii,
one only needs to prove
\eqn\eVxxxii{
\prod_{n=1}^\infty \inv{(1-q^{2n-1} ) } = \prod_{n=1}^\infty (1+q^n) . }
A simple proof is as follows:
\eqn\eVxxxiii{\eqalign{
\prod_{n=1}^\infty (1+ q^n)(1-q^{2n-1}) &= \prod_{n=1}^\infty
(1+q^{2n})(1+q^{2n-1} ) (1-q^{2n-1} ) \cr
&= \prod_{n=1}^\infty ~ (1+q^{2n}) (1-q^{4n-2}) \cr
&= \prod_{n=1}^\infty \frac{(1+q^{2n}) (1-q^{2n})}{(1-q^{4n})} ~=~ 1 . \cr}}

\def\prodd{\prod_{n=1}^\infty}

We turn now to the periodic sector.  The situation here is very different.
First note that we have already exhausted the available level $1$ highest
weight representations for the anti-periodic sector.  The true
physical vacuum resides in the periodic sector, and is characterized
by \eIIIxxv.  The momentum operators $P_1 , ~P_{-1}$, and thus the
hamiltonian must annihilate the vacuum.  Since the $\slh$ charges commute
with $P_1, ~ P_{-1}$, they must also annihilate the vacuum, otherwise
they would be spontaneously broken symmetries.  Using the expressions
in \eVviii\ specialized to the periodic sector, one indeed verifies
\eqn\eVxxxxiii{
P^L_n \rvac = Q^{\pm ,L}_n \rvac = T^L_n \rvac = 0 , ~~~~~\forall ~ n,}
and similarly for the right charges\foot{Here we have another important
distinction from the situation in conformal field theory, where, in the
latter, the vacuum is highest weight. This is possible in conformal field
theory since, after the conformal transformation $z \to \log z$, one
treats $(L-D)/2$ and $(L+D)/2$ as the right and left `hamiltonians',
and thus the above argument does not apply.}.
This implies the $\slh$ algebra must have level $k=0$ in the periodic
sector:
\eqn\eVxxxxiv{
\[ Q^{+,L}_1 , Q^{-,L}_{-1} \] ~ \rvac = \ma^2 \( T^L_0 + k/2 \)
\rvac = 0 ,~~~~~\Rightarrow ~ k=0 . }
Eq. \eVxxxxiii\ also implies
\eqn\eVxxxxv{
\[ P^L_n , P^L_m \] = 0 , ~~~~~\forall ~ n,m  }
in this sector.  The affine charges still have a meaningful
action on the other states.  For example
\eqn\eVxxxxvi{
Q^{\pm ,L}_{-1} ~ b^\mp_{-1/2} ~ \rvac = - \> b^\pm_{-3/2} ~ \rvac , }
which  says the same thing as \eIIxxxiii.

\bigskip
\newsec{Vertex Operators and Momentum Space Bosonization}
\bigskip

In this section we will describe the role played by   vertex operators.
There exists an alternative `bosonic'  organization of $\CH^{L,R}_a$.
Using the Jacobi triple
product identity\foot{One proof of this identity uses the
$\slh$ algebra\rkac.}
\eqn\eVxxxiv{
\prod_{n=1}^\infty (1-u^n v^n )(1-u^{n-1} v^n ) (1-u^n v^{n-1} )
= \sum_{m\in \Zmath} ~
(-)^m u^{m(m-1)/2} v^{m(m+1)/2} , }
one finds
\eqn\eVxxxv{
\tr{\hal} \( q^L \)  = \sum_{\al \in \Zmath + 1/2}
\frac{q^{\al^2 /2} } {\prodd (1-q^n )}  . }
This suggests the following construction.  Let $Hb^L$ ($Hb^R$) denote
the infinite Heisenberg algebra generated by $\al_n^L$ ($\al^R_n$)
$\forall ~ n$ in the anti-periodic sector (eq. \eVxvii):
\eqn\eVxxxvi{
\[ \al^L_n , \al^L_m \] = n~ \ma^{2|n|} ~ \de_{n,-m} , ~~~~~
\[ \al^R_n , \al^R_m \] = n~ \ma^{2|n|} ~ \de_{n,-m} . }
Let us define highest weight states of $Hb^{L,R}$ as satisfying
\eqn\eVxxxvii{\eqalign{
\al^L_n ~\va{\al}_L &= 0 , ~~~~~\al^L_0 ~ \va{\al}_L
= \al ~ \va{\al}_L
, ~~~~~n> 0 \cr
\al^R_n ~\va{\al}_R &= 0 , ~~~~~\al^R_0 ~ \va{\al}_R
= \al ~ \va{\al}_R ,~~~~~n>0
.\cr}}
Let us further suppose
\eqn\eVxxxviii{
L^L \va{\al}_L  = \al^2 /2 ~ \va{\al}_L , ~~~~~
L^R \va{\al}_R = - \al^2 /2 ~ \va{\al}_R  . }
Construct modules $V^L_\al$ as follows
\eqn\eVxxxix{
V^L_\al = \left\{ \al^L_{-n_1} \al^L_{-n_2} \cdots \va{\al}_L
\right\} , ~~~~~n_i > 0,}
and similarly for $V^R_\al$.  Then
\eqn\eVxxxx{
\tr{V^L_\al} \( q^L \) = \frac{q^{\al^2 /2} }{\prodd (1-q^n )}  , }
and
\eqn\eVxxxxi{
\tr{\hal} \( q^L \) = \sum_{\al \in \Zmath + 1/2} ~ \tr{V^L_\al} \( q^L \) . }
This suggests that one can identify
\eqn\eVxxxxii{
\hal = \oplus_{\al \in \Zmath + 1/2} ~~ V^L_\al . }

\def\sigh{\hat{\sigma}}
\def\phih{\hat{\phi}}
\def\alt{\tilde{\al}}

In order to construct the states $|\al \rangle_{L,R}$ explicitly,
we define some momentum space vertex
operators $\sigh^L_\al (u) $ ($\sigh^R_{\al} (u)$)
which act from $\CH^L \to \CH^L$ ($\CH^R \to \CH^R $) and transform
as follows with respect to $Hb^{L,R}$:
\eqn\eVIviii{\eqalign{
\[ \al_n^L , \sigh_\al^L (u) \] &= \al ~ \ma^{|n|} u^{-n} ~ \sigh_\al^L (u)\cr
\[ \al_n^R , \sigh_{-\al}^R (u) \] &= \al ~ \ma^{|n|} u^{n} ~ \sigh_{-\al}^R
(u)~~~~~~n\neq 0 \cr
&= -\al ~ \sigh^R_{-\al} (u) ~~~~~~~~~~~~~n=0 . \cr}}
Operators with these properties can be constructed from the
$\al_n$'s.  A requisite set of operators is given by
\eqn\eVIix{\eqalign{
\sigh^L_\al (u) &= ~ : e^{i\al \phih^L (u) } : \cr
\sigh^R_{-\al}  (u) &= ~ : e^{ i\al \phih^R (u) } : \cr}}
where
\eqn\eVIx{\eqalign{
-i \phih^L (u) &=    \sum_{n\neq 0} \ma^{-|n|} ~ \al^L_n ~ \frac{u^n}{n}
  + \al_0^L \log (u) - \alt_0^L  \cr
-i \phih^R (u) &=    \sum_{n\neq 0} \ma^{-|n|} ~ \al^R_n ~ \frac{u^{-n}}{n}
  + \al_0^R \log (u) + \alt_0^R  , \cr }}
and
\eqn\eVIxi{
\[ \al_0^L , \alt_0^L \] =
\[ \al_0^R , \alt_0^R \] = 1. }
 From \eVxi, one sees that $\phih^{L,R}$ are dimensionless operators.
In \eVIix, normal ordering is taken with respect to the
$\al_n$'s:
\eqn\eVIxii{
: \al^L_n \> \al^L_{-n} : ~ = ~ \al^L_{-n} \> \al^L_n ,
{}~~~~~~: \al^R_n \> \al^R_{-n} : ~ = ~ \al^R_{-n} \> \al^R_n ,  ~~~~~n\geq 1
.}

\def\emrvac{|\emptyset \rangle }
\def\emlvac{\langle \emptyset |}
The vacua $|\emptyset \rangle_{L,R}$ are defined to satisfy
\eqn\eVIxiii{\eqalign{
\al_n^L \emrvac_L &= \al_n^R \emrvac_R = 0 , ~~~~~n\geq 0 \cr
\alt^L_0 \emrvac_L ~&, ~~\alt^R_0 \emrvac_R \neq 0 . \cr}}
Note that $ \emrvac \equiv \emrvac_L \ot \emrvac_R $  does not
correspond to the physical vacuum since it is not annihilated by
$P_1$ and $P_{-1}$.
One has the following vacuum expectation values
\eqn\eVIxiv{\eqalign{
{}_L \emlvac ~ \phih^L (u) ~ \phih^L (u') ~\emrvac_L
&= -\log ( 1/u - 1/u' ) \cr
{}_R \emlvac ~ \phih^R (u) ~ \phih^R (u') ~\emrvac_R  &= -\log (u - u' ) \cr}}
\medskip
\eqn\eVIxv{\eqalign{
{}_L \emlvac \prod_i ~ e^{i\al_i \phih^L (u_i )} ~ \emrvac_L
&= \prod_{i< j} \( 1/u_i - 1/u_j \)^{\al_i \al_j }  \cr
{}_R \emlvac \prod_i ~ e^{i\al_i \phih^R (u_i )} ~ \emrvac_R
&= \prod_{i< j} \( u_i - u_j \)^{\al_i \al_j }  . \cr}}

Based on the structure \eIIIxii\ we define the following states
\eqn\eVIxvi{\eqalign{
\va{\al}_L ~ &= ~ \sigh^L_\al (\infty ) ~ \emrvac_L \cr
\va{\al}_R ~ &= ~ \sigh^R_\al (0 ) ~ \emrvac_R . \cr}}
Conjugate states satisfying
\eqn\eVIxvii{
{}_L \lva{-\al} \al \rangle_L
{}~=~ {}_R \langle -\al \va{\al}_R = 1 }
can also be constructed:
\eqn\eVIxviii{\eqalign{
{}_L
\lva{\al} &= \lim_{u\to 0} ~ u^{-\al^2} ~ \emlvac ~ \sigh^L_\al (u) \cr
{}_R
\lva{\al} &= \lim_{u\to \infty} ~ u^{\al^2} ~ \emlvac ~ \sigh^R_\al (u)
.\cr}}

The highest weight properties
\eVxxxvii\eVxxxviii\
are a consequence of the defining relations \eVIviii:
\eqn\eVIxx{\eqalign{
\lim_{u\to \infty} ~~ \al_n^L ~ \sigh^L_\al (u) ~ \emrvac
&= \lim_{u\to \infty} ~ ~ \al ~\ma^{|n|} ~ u^{-n} ~ \sigh^L_\al (u) ~\emrvac
\cr
&= 0 ~~~~~~~~~~n>0 \cr
&= \al ~~~~~~~~~~n=0 , \cr}}
and similarly for the right sector. Eq. \eVxxxviii\ can be established
by expressing $L^L$ ($L^R$) in terms of the operators
$\phih^L (u) $ ($\phih^R (u)$).  It is simpler to derive the result from
the vacuum expectation values
\eqn\eVIxxi{\eqalign{
\emlvac e^{-i\al \phih^L (u) } e^{i\al \phih^L (\infty)} \emrvac = u^{\al^2}
\cr
\emlvac e^{-i\al \phih^R (u) } e^{i\al \phih^R (0)} \emrvac = u^{-\al^2} .
\cr}}
Since as a differential operator in momentum space $L= u\d_u$, the Lorentz spin
of these vacuum expectation values is $\al^2 , ~ -\al^2$ respectively,
and attributing $L= \pm \al^2 /2 $ to each operator yields \eVxxxviii.

The above construction provides an exact bosonization of the
$b^\pm (u) , ~ \bb^\pm (u)$ operators in the anti-periodic sector,
if one identifies
\eqn\eVIxix{\eqalign{
b^+ (u) &=  \sqrt{\frac{\pi}{u} } : e^{i\phih^L (u)}: , ~~~~~
b^- (-u) = \sqrt{\frac{\pi}{u} } : e^{-i\phih^L (u)}: \cr
\bb^+ (u) &= \sqrt{\pi u } : e^{-i\phih^R (u)}: , ~~~~~
\bb^- (-u) = - \sqrt{\pi u} : e^{i\phih^R (u)}:  . \cr }}
One can easily check that this reproduces the 2-point functions
\eIVvii.

The above above operator formalism can easily be used to compute
the multiparticle form factors.  Following the notation of section 4,
the multiparticle form factors of the SG fields $\exp (\pm i \phi /2 )$
are defined as follows:
\eqn\effi{
f_{\pm \ha} (u_1 ,\ldots ,u_{2n} )
= \langle 0 \vert \> e^{\pm i \phi (0) /2 } \>
\vert \uparrow u_1 ; \uparrow u_2 ;\ldots \uparrow  u_n ;
\downarrow  u_{n+1} ; \ldots \downarrow  u_{2n} \rangle . }
As explained in section 4, in radial quantization these are computed
as follows:
\eqn\effii{\eqalign{
f_{\pm \ha} (u_1 ,\ldots , u_{2n} ) &=
\inv{(4\pi^2 i )^n} {}_L \langle \mp \ha \vert
b^+ (u_1 ) \cdots b^+ (u_n ) b^- (u_{n+1} ) \cdots b^- (u_{2n} )
\vert \pm \ha \rangle_L \cr
&=
\inv{(4\pi^2 i )^n} {}_R \langle \pm \ha \vert
\bb^+ (u_1 ) \cdots \bb^+ (u_n ) \bb^- (u_{n+1} ) \cdots \bb^- (u_{2n} )
\vert \mp \ha \rangle_R  . \cr }}
Using the above bosonized expressions, one finds
\eqn\effiii{\eqalign{
f_{\pm \ha} (u_1 ,\ldots , u_{2n} )
&= \inv{(4\pi i )^n }
\sqrt{u_1 \cdots u_{2n} }
\( \prod_{i=1}^n \( \frac{u_{i+n}}{u_i} \)^{\pm 1/2} \)
\( \prod_{i<j \leq n} (u_i - u_j ) \)  \cr
& ~~~\cdot \( \prod_{n+1 \leq i < j } (u_i - u_j ) \)
\( \prod_{r=1}^n \prod_{s=n+1}^{2n} \inv{u_r + u_s } \)   . \cr } }
Again these expressions agree with the known results,
though they were
originally computed using rather different methods\rmss\rform.

\bigskip
\newsec{Concluding Remarks}
\bigskip

As explained in the introduction, what began as the simple exercise
of understanding the $q\to 1$ limit of the $q-\slh$ symmetry of the
SG theory revealed a rich, previously unknown structure.  By developing
the role of undeformed affine Lie algebras in massive field theory,
we have identified the proper structures that need to be deformed
to understand massive integrable quantum field theories away from their
free points.  Though there are many important differences between the
application of affine Lie algebras to massive field theory and conformal
current algebra, many techniques developed in the context of
conformal field theory are useful in momentum space for the massive
theories.  For example, we have shown how form-factors
have virtually
the same structure as conformal correlation functions.

We have shown how the field states $\exp (\pm i \phi /2 ) |0\rangle $
are highest weight for the  level $1$ split $\slh$ symmetry.
Whether it is possible to use the affine Lie algebra to provide
a non-perturbative characterization of the
space-time correlation functions
for the fields  $\exp (\pm i \phi /2 )$
is the most important question left open by our work.

Before discussing the extension of these results to $q\neq 1$, we
remark that there are many other interesting
extensions of this work that still
involve only the undeformed affine Lie algebras since many other field
theories are known to have $q-\slh$ symmetry and they all generally
have points in the coupling where $q=1$.  We have seen that the free
fermion point of the SG theory is characterized by level $1$ affine Lie
algebra.  The algebraic structures we have introduced can be formally
extended to higher  integer level $k$, and the question arises as
to what physical models are described by these structures. We believe the
answer to this question is provided by the $k$-th fractional supersymmetric
SG theory, defined and studied in \rbl. This series of models corresponds to
the SG model $(k=1)$, super SG $(k=2)$, and in general a system of $Z_k$
parafermions interacting with a boson.  These models also have points in
the coupling constant where $q=1$.\foot{Using the definitions in \rbl, this
occurs at $\beta^2/8\pi = 1/(k(k+1))$.}  However, whereas at $k=1$
the theory can be mapped to a free one at this $q$,
for other $k$ this is not the case: inspection of the S-matrices in \rbl\
shows that the RSOS factor of the S-matrix remains non-trivial at $q=1$.
Consider also the generalization of our results to other groups.
The $\hat{G}$-affine Toda theories  have a $q-\hat{G}$ symmetry\rbl,
and also have points where $q=1$. Again inspection of the S-matrices
reveals that these are not free field theories, as the S-matrices of the
multiplets of solitons reduces to the minimal solutions (the $q-\hat{G}$
$R$-matrix pieces of the S-matrix are equal to the identity).
This discussion leads us to define the notion of the massive
$\hat{G}_k$  $q$-free quantum field theory, which is completely
characterized by a level $k$ $\hat{G}$ symmetry.  Only the
$\slh_1$ theory is genuinely free.
The terminology `q-free' refers to the fact that the  $R$-matrix
factors in the S-matrix which are characterized by $q-\hat{G}$ symmetry
are equal to the identity when $q=1$.

Since we have developed the field theory using algebraic structures,
it is relatively clear how to proceed to other values of the SG coupling
$\betah$, or beyond $q=1$, since the algebraic structures have mathematically
well-defined q-deformations.  Though many details need to be
explicitly worked out, the strategy is well-defined.
Here we outline the general aspects of
this scheme that are evident generalizations of the above results.
The analogs of the $P_n$'s are in principle known at all points
in the SG theory.  Though in \rbl\ the $q-\slh$ generators were only
constructed for the simple roots, the remaining generators
undoubtedly exist, so that the full set of $\al_n$'s exist. At the
free fermion point, we were able to construct all of these charges in
position space, but it is probably not necessary to do this in general.
A momentum space construction of the higher integrals of motion is
likely to be sufficient.  The generalization of the operators
$\bh^\pm (u)$ are evidently the Zamolodchikov-Faddeev operators
$\hat{Z}^\pm (u)$ which create asymptotic particles and satisfy
S-matrix exchange relations.  Under radial quantization these give
rise to operators $Z^\pm (u)$, $\bar{Z}^\pm (u)$, which are the
analog of the operators $b^\pm (u) , \bb^\pm (u)$.  Form-factors
should be computable as vacuum expectation values of the
$Z (u)$ operators between level $1$ highest weight states of
$q-\slh$, and here one can borrow results from \rfr,
after translating them to the principal gradation.
It should be possible to use the $\al_n$'s to
construct a momentum space bosonization as we have done at $q=1$.
Vertex operator representations of quantum affine
algebras have been developed extensively recently.  Our work
attributes a physical meaning to the `free bosons' in these
constructions as coming from the quasi-chirally split
integrals of motion $\al^{L,R}_n$ in a momentum space bosonization
scheme.

Some results presented in this paper are reminiscent of some recent
results in the context of integrable lattice models\ref\kyoto{\VVstar}.
As shown there, in infinite volume,
the finite quantum group symmetry of the lattice
hamiltonian is extended to an affine quantum group.
Exact affine quantum group symmetries  can also exist in certain
finite size chains\ref\haldane{F. D. M. Haldane, Z. N. C. Ha,
J. C. Talstra, D. Bernard and V. Pasquier, Phys. Rev. Lett 69 (1992)
2021.}.
 The quantum affine symmetry
generators act on each lattice site via a finite dimensional level $0$
representation.  It was conjectured in \kyoto\ that the Hilbert space
on lattice sites could be replaced by a level $1$ quantum affine representation
tensored with a level $-1$ representation.  It is very likely that our
results are the natural continuum version of these lattice constructions,
though a direct link has not been established since our construction
is completely
independent.  We mention some obvious parallels, and some
distinctions.
Note that if in \eVvii\ one had replaced $Q^{\pm R}_{-n} , ~
\al^R_{-n}$ with $Q^{\pm R}_n , ~\al^R_n$, then one would have obtained
a level $-1$ algebra in the right sector.
In   \kyoto\ the level $\pm 1$ modules arise when
one considers the lattice Hilbert space on two  semi-infinite lines,
one extending to the left the other to the right.  The separation
of these two spaces    defines a distinguished point at the origin.
In continuum  radial quantization, $r=0$ is a distinguished point,
and the coordinate $r$ defines a semi-infinite line.
In \kyoto\ the lattice correlation functions algebraically have
the structure of continuum form-factors, and they are expressed
as traces over quantum affine modules.  This is in contrast
to the situation in our approach, where form-factors are not
traces but instead  vacuum expectation values.
The work of \kyoto\
finds its origin in the corner transfer matrix methods, which is a lattice
analog of constructing eigenstates of the Lorentz boost operator.
The corner transfer matrix of the 8-vertex model at the free-fermion
point was studied in \rti, and this can probably be used to
relate our work with the results in \kyoto.

\bigskip\bigskip
\centerline{Acknowledgements}

I would especially like to thank D. Bernard for numerous discussions in
the early stages of this work.  I also benefited from conversations with
G. Felder, D. Haldane,  R. Rajaraman, and A. Zamolodchikov.
This work is supported
by an  Alfred P. Sloan Foundation fellowship,  and the
National Science Foundation in part through the
National Young Investigator program.

\bigskip
Note added:  After this work was completed, we learned from G. Sotkov
that the algebra \eIIxxiii{} is isomorphic to certain algebras of
conserved charges found in \ref\raass{\AASS}\ for the case of $O(2)$
invariant free massive fermion theories.

\bigskip
\appendix{A}{Twisted Affine Characters}

\bigskip

We review the aspects of affine Lie algebras needed to compute the
twisted characters \eVxxxi\rkac.

In the basis $j^a_n$ for $\slh$ \eIIxxvii, the Cartan subalgebra consists
of $j^0_0 , d$ and $k$, thus the roots are labeled by 3 numbers. A
basis of roots is provided by $\al = ({\al '} , k, n)$, $\de = (0,0,1)$,
and $\La_0 = (0,1,0)$, where $\al '$ is the root of $sl(2)$ with
${\al '}^2 = 2$.   The weights $\la$ may be expressed as
$\la = ({\la '} , k, n)$, with  $\la '$ a weight of $sl(2)$.  A basis for
$\la' $ is $\la_j ' = j \al '$, $j\in \Zmath /2$.

Define the inner product of two roots as
\eqn\eAi{
\( (\al ' , k , n) ~,~ (\beta ' , l , m ) \) = (\al' , \beta ' ) + km + n l .}
The simple roots are $\al_1 = (\al ' , 0,0)$ and $\al_0 = (-\al ' , 0 , 1)$.

Highest weights $\La$ are characterized by $\( \La , \al_i \) = n_i
\geq 0, ~~i = 0,1$. The fundamental weights are $\La_{\ha} = ( \la'_\ha,
1,0)$ and $ \La_0$.  The highest weight modules $\vh_\La$ are defined as
\eqn\eAii{
\vh_\La = \left\{ j^0_{-n_1} \cdots j^+_{-m_1} \cdots j^-_{-r_1} \cdots
\va{\La} \right\}, ~~~~~n_i , m_i \geq 1 , ~ r_i \geq 0. }

Let $[\La]$ denote the set of weights in $\vh_\La$, and define
${\rm dim} \vh_\la$ as the multiplicity of the weight $\la$. Then
the characters are defined as
\eqn\eAiii{
ch_{\vh_\La} = \sum_{\la \in [\La ]} ~ {\rm dim} \vh_\la ~ e^\la . }
Let
\eqn\eAiv{
e^\la (\hat{\al} ) = e^{(\la , \hat{\al} )} , }
where $\hat{\al}$ parameterizes an arbitrary root:
\eqn\eAv{
\hat{\al} = - 2 \pi i ( z\al ' + \tau \La_0 + \mu \de ) . }
Then the evaluated characters are defined as
\eqn\eIvi{
ch_{\vh_{\La}} (z , \tau, \mu ) \equiv ch_{\vh_\La} ~ (\hat{\al} ) . }
For level $k=1$, the integrable
highest weight states are $\La_0 , \La_{1/2}$, and
the characters are explicitly given by the formula
\eqn\eAvii{
ch_{\vh_{\La_j}} (z ,\tau , \mu ) = e^{-2\pi i \mu } ~~
\frac{\sum_{n\in \Zmath} ~ q^{n(n+2j)} ~ e^{-4\pi i z (n+j )} }
{\prod_{m \geq 1} (1-q^{n} )} , }
where
$$q \equiv e^{2\pi i \tau }. $$

In order to relate this result to the traces used in section 5, note
that
\eqn\eAviii{
ch_{\vh_{\La}} (z , \tau , \mu = 0 ) = \tr{{\vh_{\La}}} ~
\( q^{-d} ~ e^{-2\pi i \> z \> j^0_0} \) . }
Recalling that in the twisted representation \eIIxxx\
$d' = 2d + j^0_0 /2 $, one finds
\eqn\eAix{
\hat{\chi}^{tw}_{j} \equiv \tr{\vh_{\La_j}} \( q^{-d' } \)
= ch_{\vh_{\La_j}} ( z= \tau/2 , \tau^2 , \mu=0 ) . }

It is useful to derive a different expression for the same twisted
characters.  The characters $\chih^{tw}_{j}$ are precisely
the so-called principally specialized characters.  The
derivation $d'$ satisfies $(\al_i , d' ) = 1$ for $\al_i$ a simple
root.
Define
\eqn\eAx{
F_1 (e^\la ) = q^{(\la , d' )} .  }
Then
\eqn\eAxi{
\tr{\vh_\La} ~ \( q^{-d'} \) = F_1 \( e^{\La } ch_{\vh_\La} \) . }
For level $1$ these were computed in \rkac:
\eqn\eAxii{
\chih^{tw}_0 = \chih^{tw}_{\ha} = \prod_{n\geq 1} ~
\inv{ (1-q^{2n-1} ) } .  }

\listrefs
\end